\documentclass[5p, authoryear]{elsarticle}
\usepackage{natbib, amsmath, amssymb, graphicx, lineno, multirow, ulem, array, cancel, url, aas_macros}
\usepackage[usenames]{xcolor}

\newcommand{\gr}{$\gamma$-ray}
\newcommand{\grs}{$\gamma$-rays}

\newcommand{\revision}[2]{#2}

\begin{document}

\title{An image-based array trigger for Imaging Atmospheric Cherenkov Telescope Arrays}
\author[isu,umn]{Hugh Dickinson\corref{cor1}}
\ead{hdickins@umn.edu, hughd@iastate.edu}
\author[isu]{Frank Krennrich}
\ead{krennrich@iastate.edu}
\author[isu]{Amanda Weinstein}
\ead{amandajw@iastate.edu}
\author[isu]{Jonathan Eisch}
\ead{jeisch@iastate.edu}
\author[anl]{Karen Byrum}
\ead{byrum@anl.gov}
\author[anl]{John Anderson}
\ead{jta@anl.gov}
\author[anl]{Gary Drake}
\ead{jta@anl.gov}

\address[isu]{Department of Physics \& Astronomy, Iowa State University, Zaffarano Hall, 2334 Pammel Drive, Ames, IA 50011-3160, United States of America}
\address[umn]{School of Physics and Astronomy, 275-04 Tate Laboratory of Physics, 116 Church St. SE, Minneapolis, MN 55455, United States of America}
\address[anl]{High Energy Physics Division, 9700 S. Cass Avenue, Building 362, Argonne, IL 60439, United States of America}

\date{}

\cortext[cor1]{Corresponding author}

\begin{abstract}
 It is anticipated that forthcoming, next generation, atmospheric Cherenkov telescope arrays will include a number of medium-sized telescopes that are constructed using a dual-mirror Schwarzschild-Couder configuration. These telescopes will sample a wide ($8^{\circ}$) field of view using a densely pixelated camera comprising over $10^{4}$ individual readout channels. A readout frequency congruent with the expected single-telescope trigger rates would result in substantial data rates. To ameliorate these data rates, a novel, hardware-level Distributed Intelligent Array Trigger (DIAT) is envisioned. A copy of the DIAT operates autonomously at each telescope and uses reduced resolution imaging data from a limited subset of nearby telescopes to veto events prior to camera readout {and any subsequent network transmission of camera data that is required for centralized storage or aggregation}. We present the results of Monte-Carlo simulations that evaluate the efficacy of a ``Parallax width'' discriminator that can be used by the DIAT to efficiently distinguish between genuine gamma-ray initiated events and unwanted background events that are initiated by hadronic cosmic rays.
\end{abstract}

\begin{keyword}
 Gamma-Ray Astronomy \sep Array Trigger \sep Gamma-Hadron Separation \sep Advanced Algorithm \sep Energy Threshold Reduction
\end{keyword}

\maketitle

\section{Introduction}\label{sec:intro}


{Gamma-ray astronomy with ground-based instruments was enabled by the pioneering efforts by the Whipple collaboration \citep{1989ApJ...342..379W} by demonstrating the excellent inherent sensitivity of imaging atmospheric Cherenkov telescopes (IACTs) for detecting  astrophysical TeV photons.}

{The detection of several astrophysical sources (the Crab Nebula and several blazars including Mrk 421, Mrk 501, 1ES 2344+524, H 1426+428, 1ES 1959+650) with the Whipple 10~m, HEGRA and CAT instruments in the mid-1990s motivated construction of the current generation of telescopes (H.E.S.S. \citep{2006A&A...457..899A}, MAGIC-II \citep{2012APh....35..435A} and VERITAS \citep{2006APh....25..391H}) in the early twenty first century.  These modern instruments employ the stereoscopic imaging technique and exceed the sensitivity of any previous TeV gamma-ray telescope by an order of magnitude. The future potential of TeV astronomy to explore high energy astrophysics and particle astrophysics is amply demonstrated by a collective catalog of over two hundred TeV $\gamma$-ray sources\footnote{TeVCat: \url{http://tevcat.uchicago.edu}}.}

{Observations of celestial photons with energies in the interval 20 GeV - \revision{100}{300} TeV address important questions of modern astrophysics and particle physics. Indeed, \gr\ astronomy promises key insights for a diverse range of topics including: the origin of cosmic rays; particle acceleration and propagation; cosmological radiation and magnetic fields and even the composition and nature of dark matter \citep[see e.g.][]{Acharya20133}. Future IACTs will also provide important follow-up observations of sources that are identified at energies above 0.1 GeV by the \revision{the}{} \textit{Fermi} space telescope's all-sky survey. To date, \textit{Fermi} has delivered a catalog containing over 3000 distinct \gr\ sources \citep{2015ApJS..218...23A}, of which only a small fraction have been studied at multi-TeV energies. The forthcoming Cherenkov Telescope Array \citep[CTA;][]{2013APh....43....3A} will extend the energy coverage of the \textit{Fermi} survey to span six orders of magnitude in energy, with sufficient sensitivity to perform statistically rich population studies for source classes in the 100~TeV regime. CTA will exhibit substantially improved angular resolution and reduced background contamination with respect to current-generation instruments. These enhancements \revision{will permit}{are expected to permit the establishment of} mutiwavelenth associations \revision{to be established}{} for over 1000 sources that were discovered by Fermi but lack a plausible counterpart at other wavelengths.}

{Building on the nascent successes of this new research field, the CTA project has been undertaken by a world-wide collaboration of scientists who have coalesced around the shared goal of constructing and operating \revision{a single next-generation IACT}{a next-generation IACT array}. The basic concept of CTA \citep[see e.g.][]{Acharya20133} envisions heterogeneous arrays comprising telescopes with differing sizes and capabilities. The prevalent designs include a large number of \revision{small-size telescope}{small-sized telescopes} (SSTs, $\rm \varnothing_{mirror} \sim 4~m$), several tens of medium-size telescopes (MSTs, $\rm \varnothing_{mirror} \sim 10 - 12~m$) and a small number of large-size telescopes (LSTs, $\rm \varnothing_{mirror} \sim 23~m$), which combine to provide wide energy coverage spanning 20~GeV - 100~TeV.}

{The MSTs probe the sub-TeV to multi-TeV energy band, which is \revision{}{a} regime for which the IACT technique achieves maximal sensitivity and excellent angular resolution. At higher energies, the Cherenkov light intensity for \gr\ showers is subtantially increased, such that detection and imaging becomes feasible using smaller mirror areas provided by SSTs. Conversely, large mirror areas provided by the LSTs are required in order to access the 20 - 100 GeV regime, for which the Cherenkov light intensity is faint.}

{The energy ranges to which each telescope type is sensitive exhibit substantial overlap but the effective collection areas \citep[see e.g.][]{Acharya20133} for sub-arrays comprising the MSTs and LSTs are markedly disparate ($\rm \sim 1 km^{2}$ and $\rm \sim 0.1 km^{2}$, respectively). A substantial increase in low-energy event statistics could be achieved by enabling detection of $\lesssim100\;\mathrm{GeV}$ GeV \grs\ using the MST sub-array, thereby augmenting the much smaller effective collection area of the LSTs in this energy regime. Accordingly, it is important to consider how improvements in their electronics designs could enhance the low energy response of MSTs.}

{Compelling scientific motivation for improving the sensitivity in the sub-100~GeV regime is provided by a renewed interest in \gr\ emission from pulsars, especially examples that exhibit an unexpected emission component in the 10 - 100 GeV regime. A further motivation is the potential to expand the size of the observable Universe for IACTs. On cosmological distance scales, the extragalactic background light strongly attenuates photons with energies exceeding $\sim10$s of GeV according to an opacity that increases with \gr\ energy \citep[e.g.][]{Dwek2013112}.}

{In this paper we describe a \textit{Decentralized Intelligent Array Trigger} (DIAT) system that is specifically designed to enable opera\revision{ta}{}tion of IACTs with a low energy threshold, while maintaining stable and manageable data rates in the presence of varying observing conditions and ambient illuminaton. Transient brightening of the night sky background light (NSB) can occur for several reasons including partial cloud coverage during moonless nights, observing the bright regions in the galactic plane or observation with partial moonlight. Stable telescope operation across these regimes would require the adjustment of the \textit{single telescope} trigger threshold, if the rates were not moderated by a hardware \textit{array} trigger. Telescope and array triggering strategies are discussed further in $\S$\ref{sec:tel_trig}.}

{In addition to mitigating the effect of bright NSB, the DIAT system we present is also capable of substantially reducing the background trigger rate produced by cosmic-ray-induced air showers. Configurable firmware can selectively tune the background rate suppression from factors of a few up to two orders of magnitude, while maintaining an acceptably high acceptance for $\rm gamma$-ray events.}

{The capabilities of the DIAT system are particularly well suited for a highly innovative telescope design \citep[e.g.][]{2008ICRC....3.1445V}, which uses a dual-mirror \textit{Schwarzschild Couder} (SC) configuration, and incorporates a finely pixelated camera comprising 11,328 independent silicon photomultiplier-based readout channels \citep{2015arXiv150902345O}.}

{This SC telescope (SCT) design provides a wide field of view ($\sim8^{\circ}$) and high-resolution imaging of air showers with excellent angular and energy resolution. However, the large number of pixels combined with a nominal camera trigger rate of 10~kHz per telescope implies a substantial cost in terms of data transfer, storage and processing requirements. Reduction in overall data volume can be achieved at the trigger level using front-end electronics and potentially further by off-line post-processing.}




\section{Telescope and Array-Level Triggering}\label{sec:tel_trig}

IACTs operate in a strongly background-dominated regime. \textit{Array} triggering schemes use information from multiple telescopes to veto many background events \textit{before} camera readout, with the goal of stabilizing the array's energy threshold and dead time under variable ambient illumination. For the densely pixelated SCT camera, control over the array trigger rate is also desirable to guarantee that data-transfer rates remain tractable at the extremes of normal observing conditions.

{The overwhelming majority of individual \textit{pixel} triggers are associated with low-level illumination from the ambient night-sky background \revision{(NSB)}{} light. The NSB induces a rate of random pixel triggers that is often sufficient to generate a large rate of spurious individual \textit{telescope} triggers.}

{{To address this issue, current-generation IACTs implement \textit{multi-level} hardware trigger systems that require sequential fulfillment of criteria that involve signals from an increasing number of imaging elements \citep[See e.g.][for more details regarding the trigger systems of H.E.S.S., MAGIC-II and VERITAS]{2004APh....22..285F,2011ITNS...58.1685M,2008ICRC....3.1539W,2013arXiv1307.8360Z,2007ITNS...54..404P,2016JInst..11P4005L}}. Triggering of a single \textit{telescope} typically requires a cluster of adjacent camera pixels to trigger within a temporal coincidence window lasting a few nanoseconds. This precaution subtantially reduces the rate of telescope triggers caused by the accidental pileup of night sky background photons in a single pixel.}

{For typical sky brightnesses, the intensity of NSB photons is such that the rate of NSB-induced triggers becomes negligible for per-pixel signals exceeding $\sim5$ photons.} For larger {per-pixel} photon intensities, another background component comprising \textit{temporally correlated} Cherenkov light from cosmic-ray (CR) initiated air showers becomes dominant.
{Modern multi-telescope IACT arrays reduce contamination from CRs using some variant of a \textit{multiplicity} trigger that requires a minimum of two telescopes to trigger within 50 - 100~ns. Basic multiplicity array trigger systems help to stabilize the data rates at which these telescopes must operate, primarily by rejecting a faint subset of cosmic-ray- and single-muon-induced showers that trigger only a single telescope. CR-initiated air-showers that trigger multiple telescopes often exhibit temporal correlation between individual telescope triggers that is sufficient to impair the efficacy of traditional multiplicity array triggers.}

{The DIAT concept represents a significant extension to the functionality of existing array trigger schemes. It is the first system to use imaging information to discriminate between \gr- and cosmic-ray-induced shower images in real time at the \textit{hardware trigger} level. This has become possible possible thanks to the recent emergence of fast Field Programmable Gate Arrays (FPGAs) that can evaluate and apply individual camera trigger criteria while simultaneously performing reduction of the camera pixel hit pattern into summary image parameters in real-time. Rapid computation using the image parameters that are generated by neighbouring telescopes within the array permits near-real-time stereo analysis of the event.}

Subsequent sections demonstrate that spurious \revision{triggers CR-induced}{CR-induced triggers} \textit{can} be identified and vetoed in near-real-time by processing reduced image data using modern FPGAs. \revision{W}{In section \ref{sec:pwidth_def} we define the \textit{parallax width} event discriminator and mathematically describe the algorithm that is used to compute it. Section \ref{sec:passthrough} outlines how excessive trigger suppression for high energy \gr\ events can be ameliorated by indiscriminately accepting events that surpass a combined image brightness threshold. In section \ref{sec:hw_implementation}, we provide a brief description of a distributed hardware array trigger that could be used to deployed to compute the \textit{parallax width} in near real-time for large IACT arrays. Section \ref{sec:results} presents and examines our results, which were obtained by applying the \textit{parallax width} algorithm to discriminate between simulated \gr- and proton-like events for a variety of \gr\ source configurations. We summarize our results in section \ref{sec:conclusions}. Overall, w}e explicitly show that image parameters involving only the first moments of each telescope image \revision{image}{} are sufficient to discriminate between hadronic and \gr-induced showers. If the rate of spurious telescope triggers can be effectively controlled, then the sensitivity of an IACT array to low energy \grs\ can be enhanced by reducing the trigger threshold of individual camera pixels.

{We note that there exist alternative proposals for data rate suppression that do not use multi-telescope information, and do not veto events in their entirety. Instead, such schemes implement strategies for lossy compression of event data by discarding segments of the Cherenkov image that are signal-free or deemed unlikely to contain signals produced by Cherenkov light \citep[e.g.][]{2015arXiv150807584C,2013JInst...8P6011N,2016JInst..11P4005L}. }


\section{The Parallax Width Discriminator}\label{sec:pwidth_def}

Electromagnetic air-showers initiated by \grs\ are characterized by a single shower axis, and triggered telescopes capture coherent elliptical images with a well defined nucleus-within-coma structure. In contrast, the hadronic component of cosmic-ray showers produces multiple sub-showers and typically yields a more fragmentary distribution of Cherenkov light \citep[see e.g.][]{1997JPhG...23.1013F}. The \textit{parallax width} \citep{1995ExpAstro...6...285,10.1063/1.3076821,2009arXiv0908.0179S} discriminator ($P$) leverages the difference between CR and \gr\ shower images to rapidly distinguish between these event categories at the \textit{hardware} level. It is instructive to separate the distinct algorithmic stages that comprise the computation of $P$ into {two separate categories: Camera Image Preprocessing, and Computation of the Parallax Width Discriminator which are described in $\S$\ref{sec:im_preproc} and $\S$\ref{sec:pwidth_comp}, respectively.}

{\subsection{Telescope Simulation Framework}\label{sec:simulations}
 To investigate the efficacy of $P$ to distinguish \gr\- and cosmic-ray initiated air-showers, simulations of the SCT array configuration 
 were produced using the \texttt{sim\_telarray} software package \citep{2008APh....30..149B}. Simulations of  \gr\ emission from point-like and diffuse \footnote{\revision{}{\gr\ events were simulated with randomly distributed arrival directions, distributed within a cone with a $10^{\circ}$ opening angle around the telescope pointing axis. Such an event distribution appears genuinely diffuse for the SCT, which has an 8 degree field of view.}} sources, as well as simulations of a diffuse proton background were used. The energy spectra of the simulated \grs\ and protons described falling power-law distributions ($dN/dE\propto E^{-\Gamma_{x}}\;:x\in\{\gamma, p\}$) with spectral indices $\Gamma_{\gamma}=2$ and $\Gamma_{p}=2.7$, respectively\footnote{\revision{}{Although \gr\ spectra with indices as hard as 2 are seldom observed in nature, using simulated events with $\Gamma_{\gamma}=2$ enables the performance of the parallax width trigger to be evaluated with good number statistics for high \gr\ energies.}}. Unless it is stated otherwise, \gr\ events model astrophysical sources located centrally in the field of view at an altitude of $70^{\circ}$ and an azimuth of $180^{\circ}$. The raw simulated camera images were provided as input to a computer simulation of the DIAT array triggering scheme.}

\subsection{Camera Image Preprocessing}\label{sec:im_preproc}

\begin{figure*}
 \centering
 \includegraphics[width=0.5\textwidth]{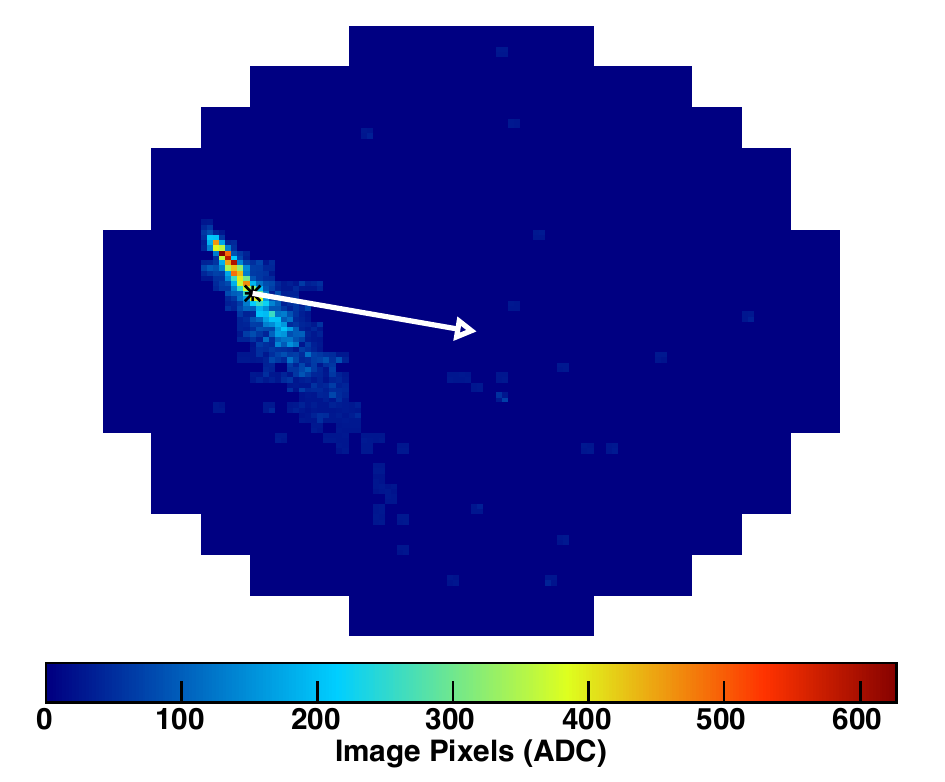}
 \includegraphics[width=0.5\textwidth]{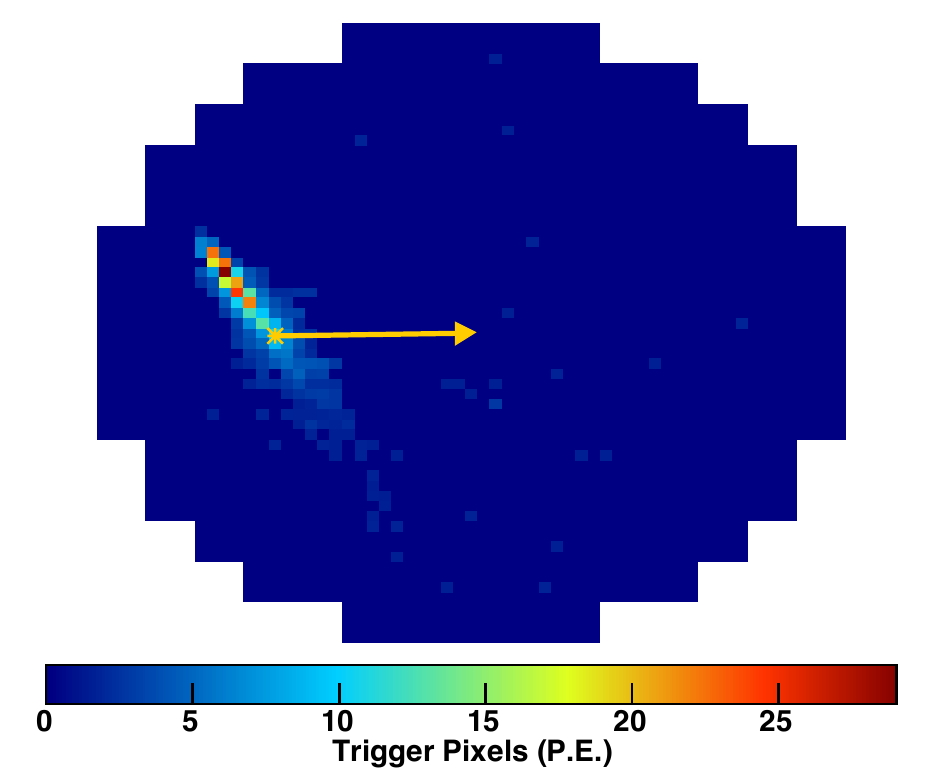}\\
 \includegraphics[width=0.5\textwidth]{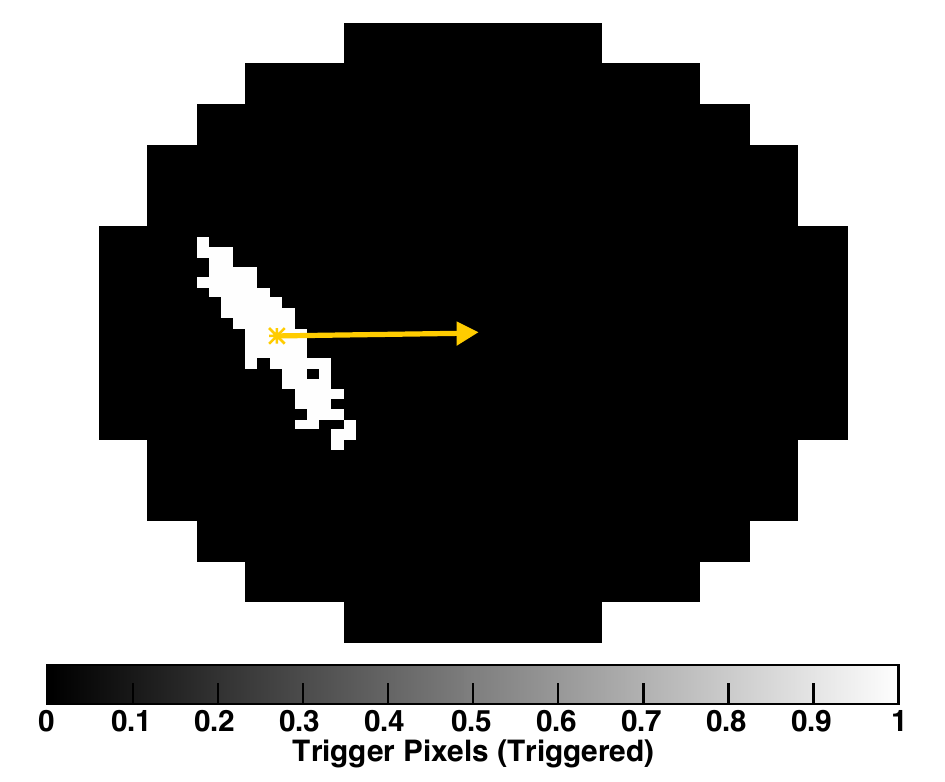}
 \caption{\small The \textit{upper-left-hand} panel illustrates an unprocessed, finely pixelated image corresponding to a \gr-initiated air-shower.  The \textit{upper-right-hand} panel illustrates the intermediate, coarsely pixelated image that is derived by combining the signals from sets of 4 neighbouring imaging pixels. The \textit{lower} panel illustrates the Boolean-valued trigger image that is required by the algorithm that computes $P$. The \textit{orange} arrows correspond to the vector $\mathbf{r}_{F}$, which connects the implicitly \textit{unweighted} trigger-image centroid with the \textit{fiducial} camera-plane coordinate $\mathbf{r}^{\star} = (0,0)$ at the camera centre. For comparison, the \textit{white} arrow  (\textit{top-left-hand} panel) connects the \textit{signal-amplitude-weighted} mean position of all \textit{imaging} pixels with $\mathbf{r}^{\star}$.}\label{fig:P_computation_Camera}
\end{figure*}

Before $P$ is computed, preprocessing algorithms are applied to each camera image at the single telescope level. Figure \ref{fig:P_computation_Camera} illustrates the manner in which telescope-specific camera data are processed to derive a compact geometrical representation of the captured shower image.

To mitigate the effect of Night Sky Background
photons randomly triggering each of the 11,328 individual readout channels, and thereby enable a reduction of the trigger threshold, 2,832 lower resolution \textit{super-pixels}\footnote{{Throughout this work the  terms \textit{super-pixel} and \textit{trigger pixel} are considered to be synonymous and are used interchangeably.}} are formed from four adjacent imaging pixels. The combined signal amplitudes of the super-pixels are used to form a Boolean-valued \textit{trigger image} with a predefined threshold on the summed output level ($n_{\rm pe}$, typically expressed in photoelectron-equivalent units, hereafter \textit{p.e.}) segregating true (triggered) and false values.  An \textit{individual telescope} is deemed to have triggered if its Boolean trigger image includes 3 or more adjacent triggered super-pixels\footnote{\revision{}{Super-pixels are considered to be adjacent if they share at least one common vertex.}}.

\begin{figure*}[p]
 \centering
 \includegraphics[width=0.5\textwidth]{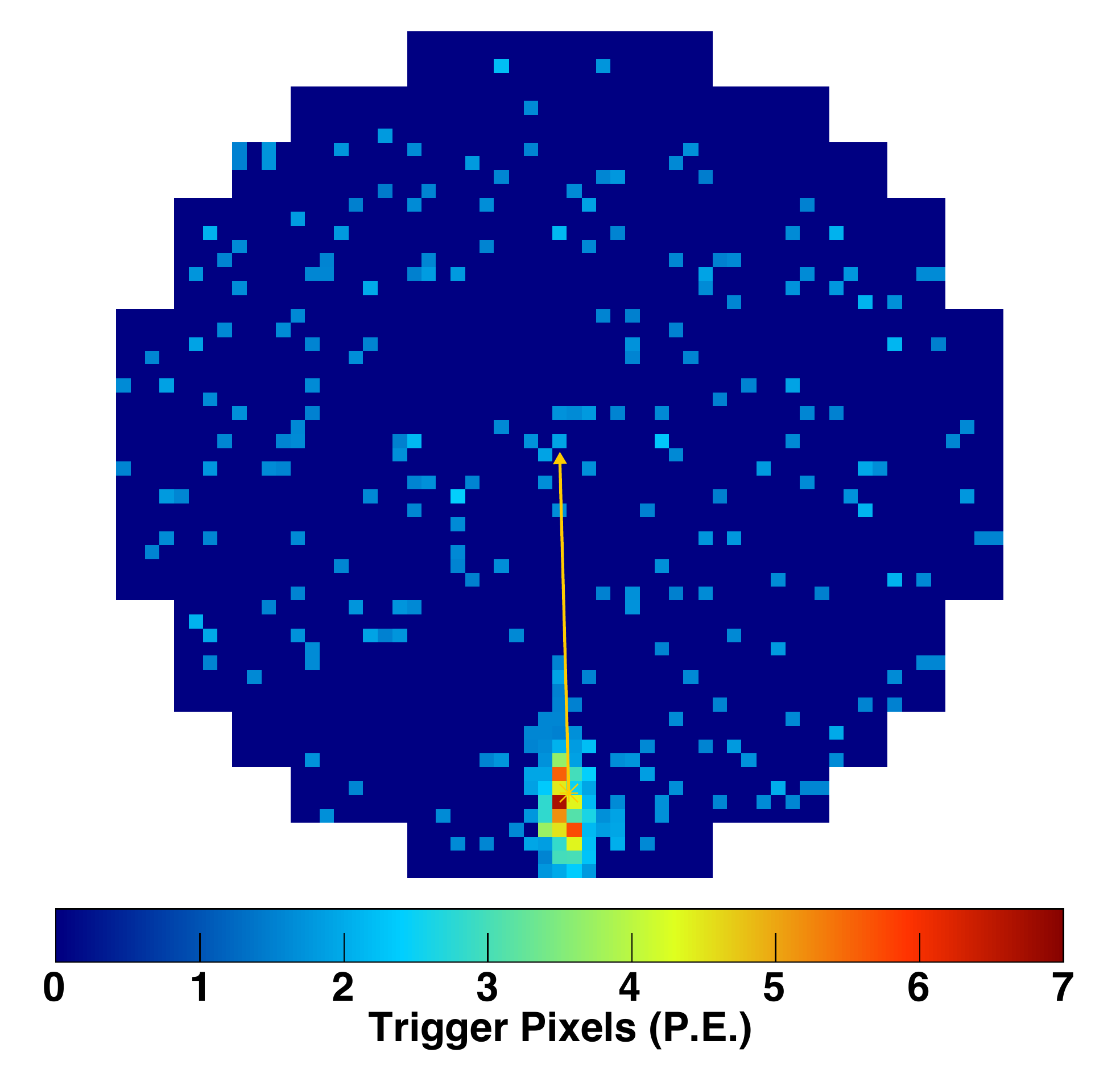}
 \includegraphics[width=0.5\textwidth]{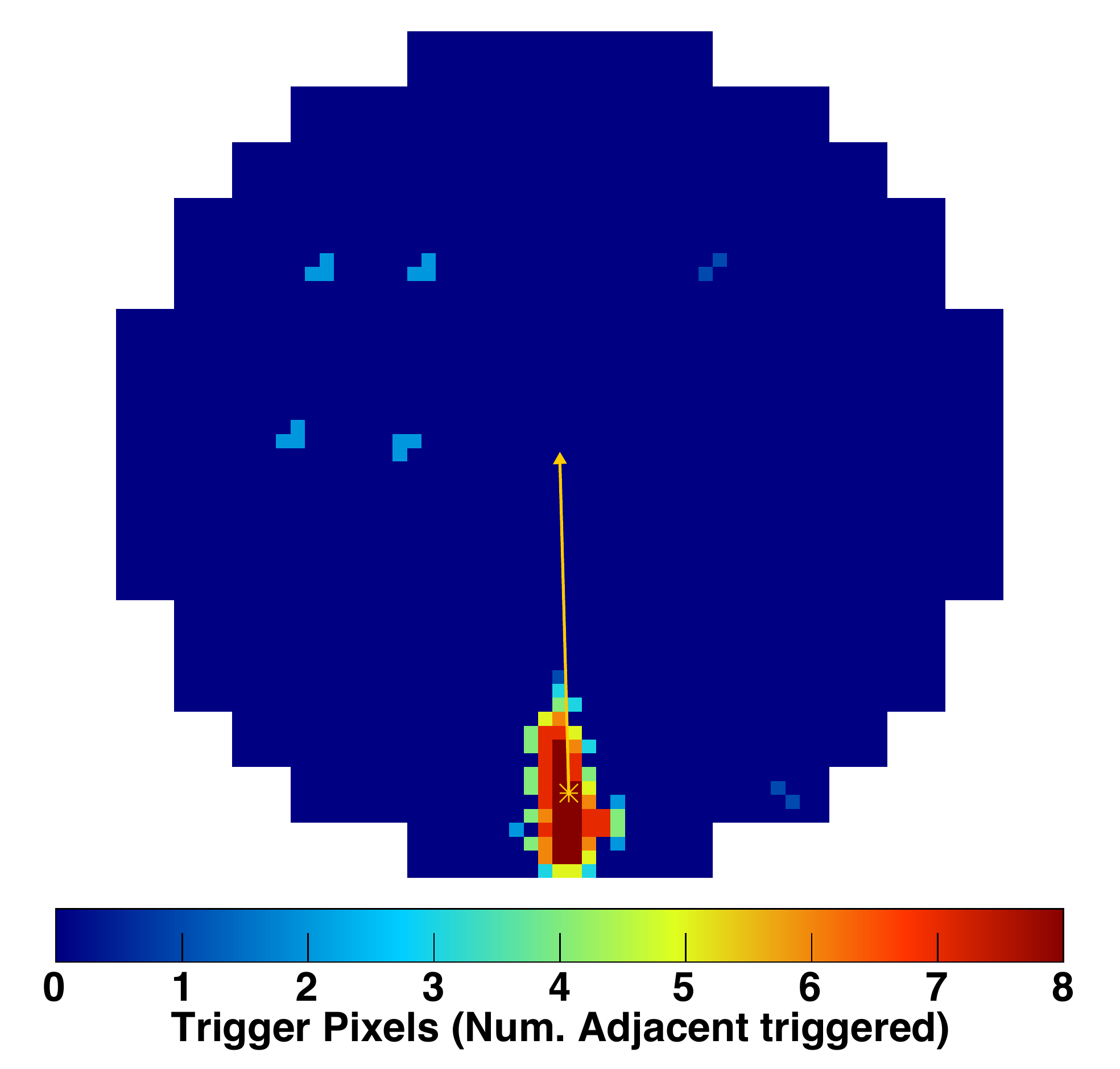}\\
 \includegraphics[width=0.5\textwidth]{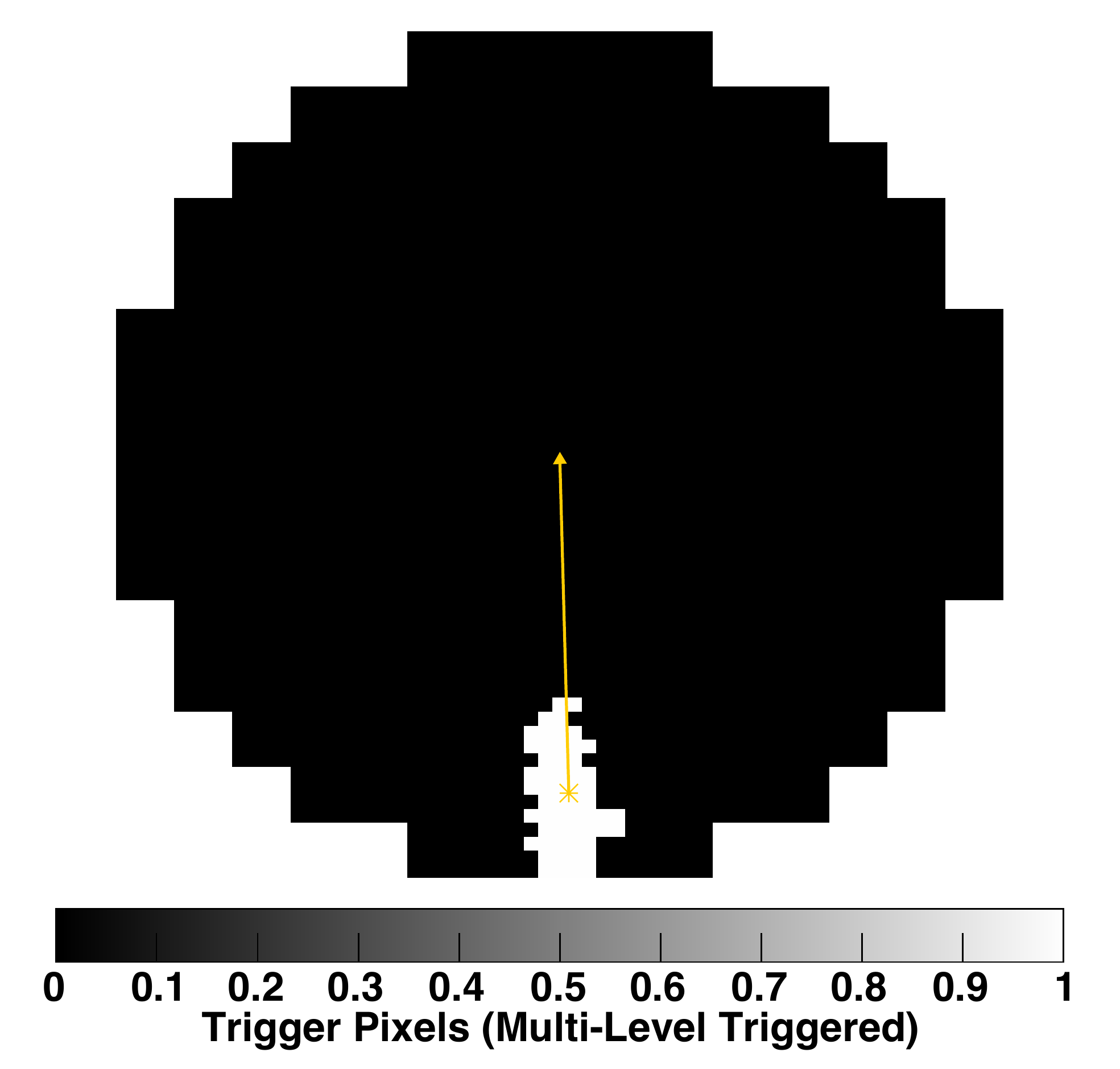}
 \caption{\small Demonstration of the two-level cleaning algorithm for $n_{1} = 3$, $n_{2} = 5$ and a pixel threshold $n_{\rm pe} = 2\,{\rm p.e.}$ when applied to a genuine \gr\ image. The \textit{upper-left} panel shows signals registered by each triggered superpixel in photoelectron equivalent counts. The number of triggered neighbours for each super-pixel is illustrated in the \textit{upper-right} panel. The \textit{lower} panel reveals the result of applying the two-level cleaning algorithm  to the original trigger image.}\label{fig:two_level_cleaning_gamma}
\end{figure*}

\begin{figure*}[p]
 \centering
 \includegraphics[width=0.5\textwidth]{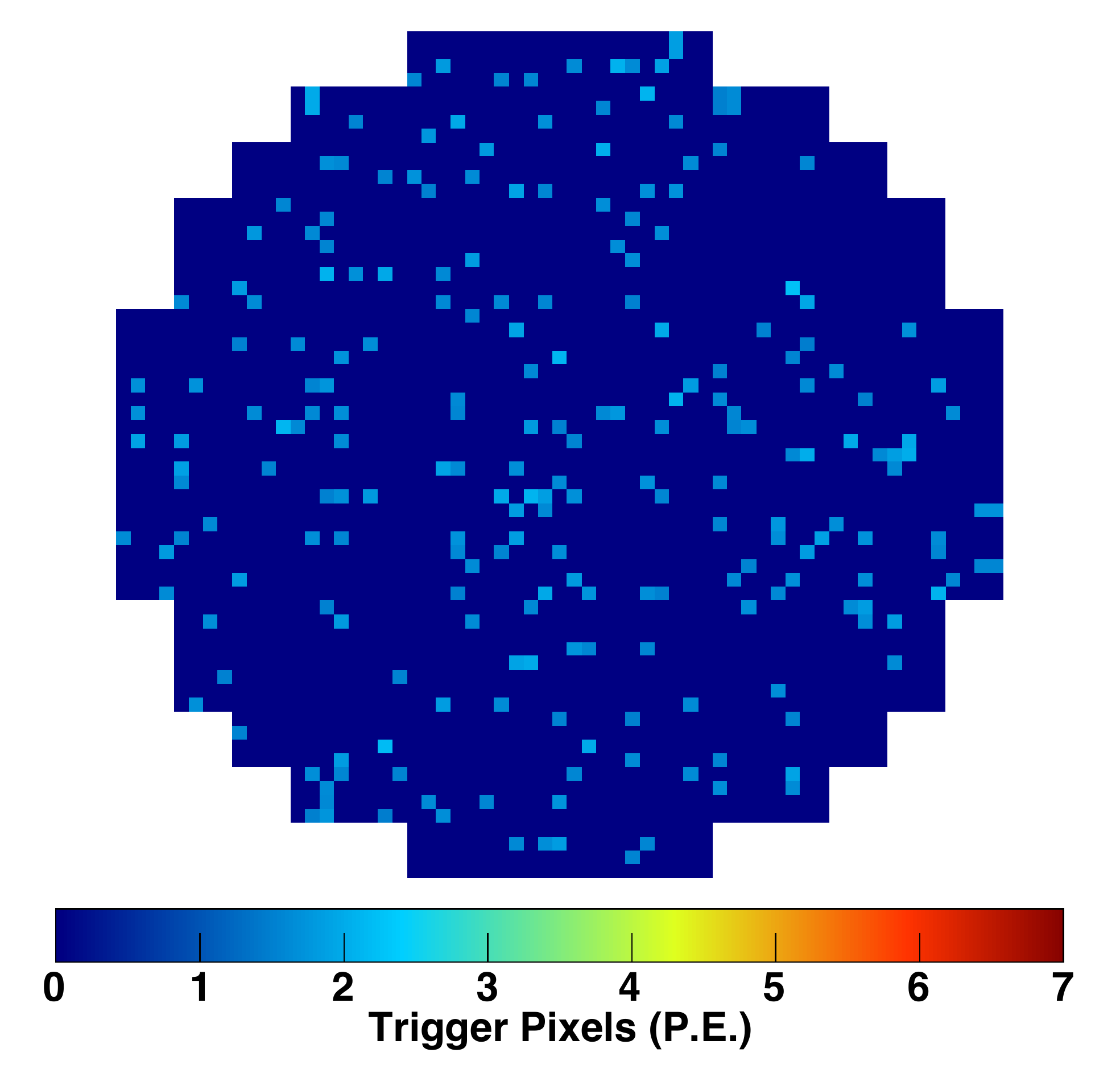}
 \includegraphics[width=0.5\textwidth]{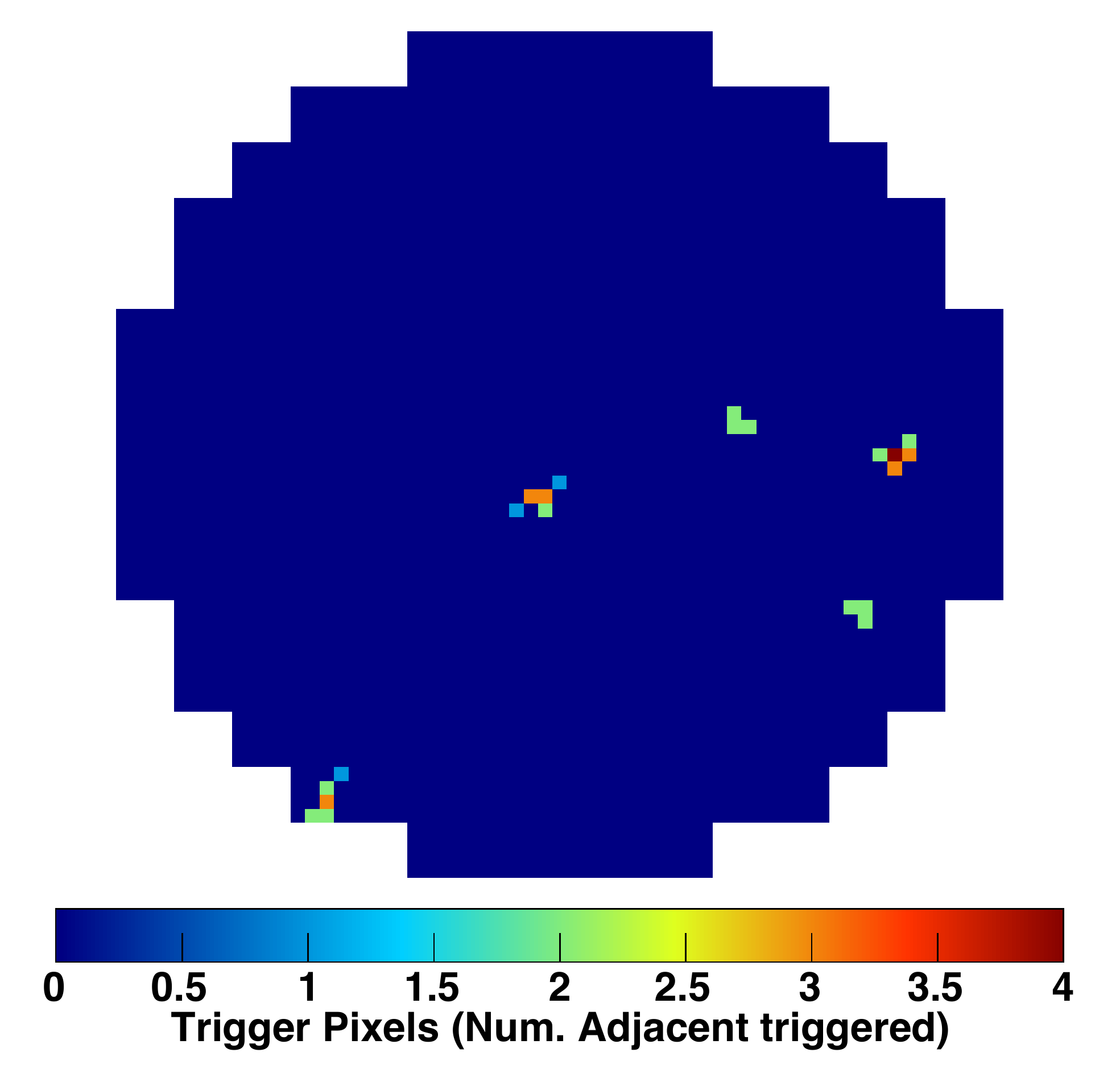}\\
 \includegraphics[width=0.5\textwidth]{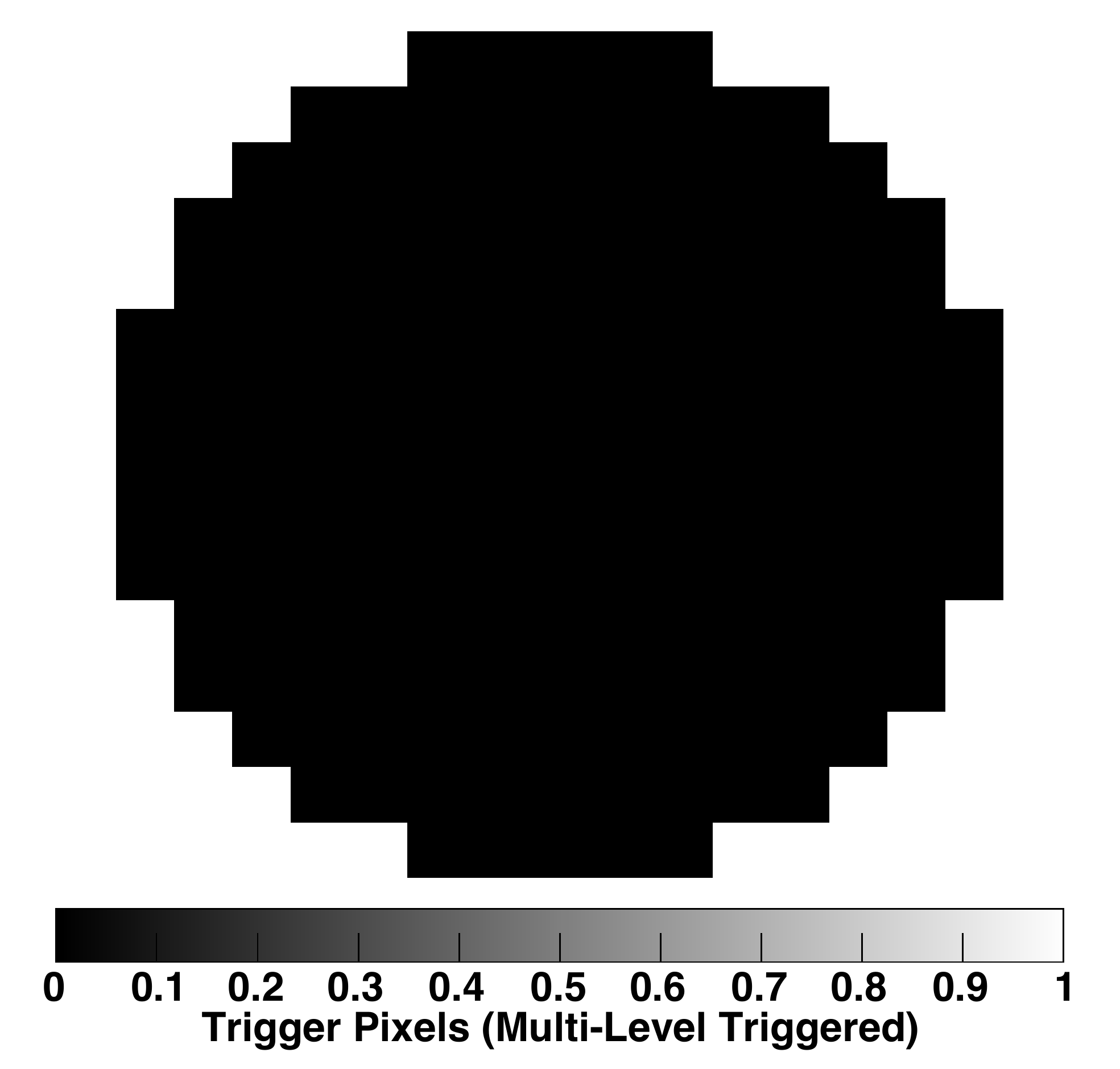}
 \caption{\small Demonstration of the two-level cleaning algorithm for $n_{1} = 3$, $n_{2} = 5$ and a pixel threshold $n_{\rm pe} = 2\,{\rm p.e.}$ when applied to an image comprising randomly triggered super-pixels that would otherwise be sufficiently numerous to fulfill the pass-through criterion ($n_{\rm TP} > 16$). The panels correspond to their counterparts in Figure \ref{fig:two_level_cleaning_gamma}.}\label{fig:two_level_cleaning_noisy}
\end{figure*}

A two-level image cleaning algorithm is applied to each trigger image to remove small clusters of NSB-induced pixel triggers that can incorrectly trigger the telescope or bias subsequent computation of $P$ by contaminating valid telescope images with noise. Nominally triggered pixels are retained in the cleaned trigger image if valid triggers were generated by \textit{at least} $n_{1}$ immediately adjacent pixels, \textit{at least one} of which \textit{itself} has $n_{2}$ triggered neighbours\footnote{The multilevel neighbour multiplicities for the cleaning algorithm were fixed to $n_{1}=3$ and $n_{2}=5$ after empirical investigation indicated good performance using these values.}. This cleaning algorithm is designed to retain extended contiguous groups of trigger pixels that are consistent with genuine air-shower images. Figures \ref{fig:two_level_cleaning_gamma} and \ref{fig:two_level_cleaning_noisy} demonstrate the algorithm for an image of a \gr-initiated air shower event and an image that contains 
a large number of randomly triggered super-pixels, respectively. Most triggered pixels in the genuine \gr\ event are retained, while the sparse pixels that are randomly triggered are removed.

\subsection{Computation of the Parallax Width Discriminator}\label{sec:pwidth_comp}

\begin{table*}
 \begin{tabular}{lcm{0.7\textwidth}}
  \hline
  Coordinate System & Symbol                           & Description                                                                                                                                        \\
  \hline
  \multirow{3}{*}{Camera Plane}
                    & $\mathbf{r}_{C,i}$               & The coordinates of the centroid of the boolean-valued \textit{trigger image} for the $i$th telescope.                                              \\
                    & $\mathbf{r}^{\star}_{F,i}$       & The vector pointing \textbf{from} the centroid of the boolean-valued \textit{trigger image} \textbf{to} the camera centre for the $i$th telescope. \\
  \hline
  \multirow{3}{*}{Mirror Plane}
                    & $\mathbf{r}^{\prime}_{F,i}$      & The projection of $\mathbf{r}^{\star}_{F,i}$ \textbf{from} the camera plane \textbf{into} the array mirror plane for each telescope.               \\
                    & $\mathbf{r}_{\times,j}^{\prime}$ & The mirror plane coordinates of a forward intersection between the $\mathbf{r}^{\prime}_{F,i}$ for two separate telescopes.                        \\
  \hline
 \end{tabular}\caption{Notation used in the derivation of $P$.}\label{tab:p_deriv_vars}
\end{table*}

\begin{figure*}[p]
 \center
 \includegraphics[width=0.68\textwidth]{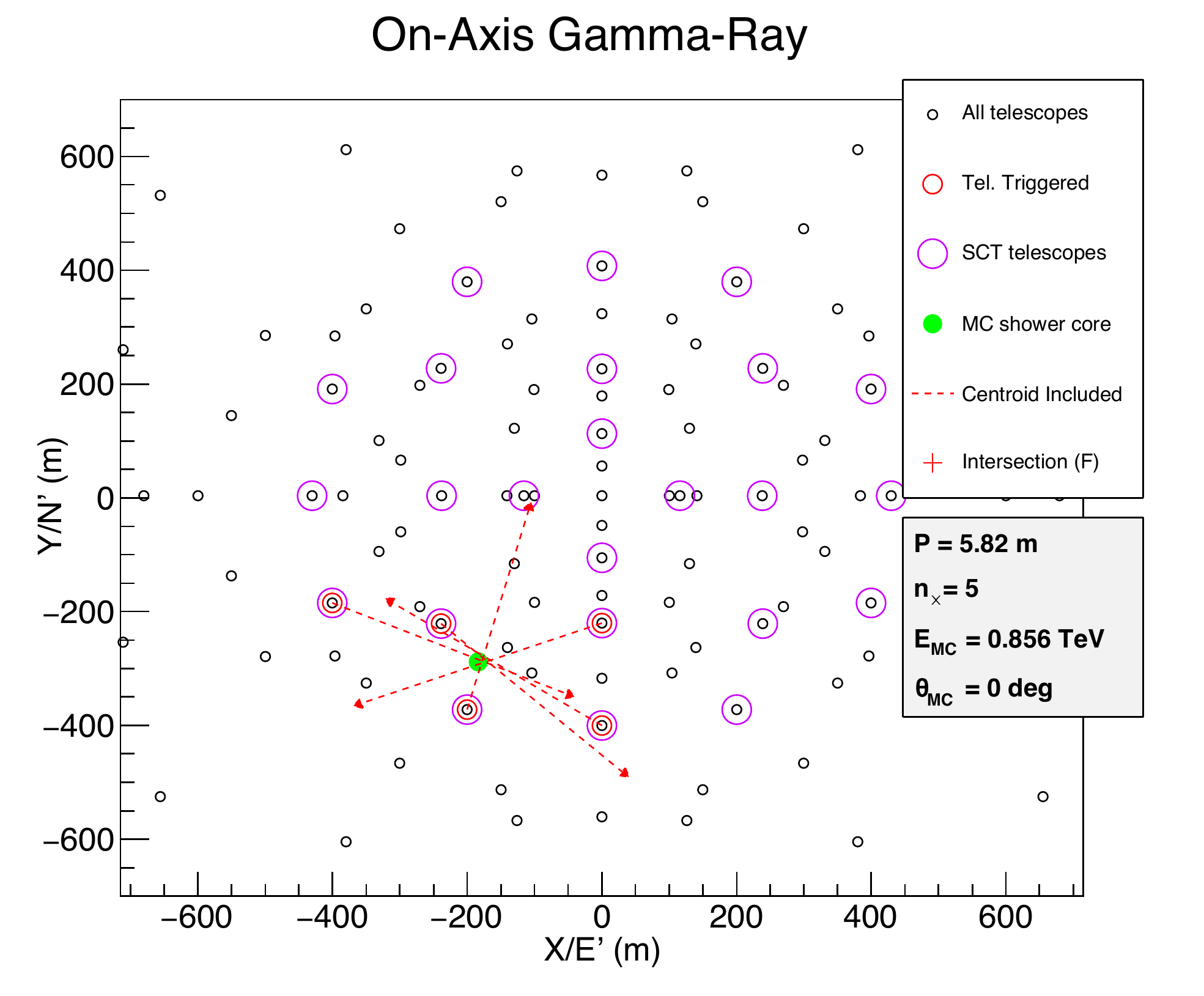}\\
 \includegraphics[width=0.68\textwidth]{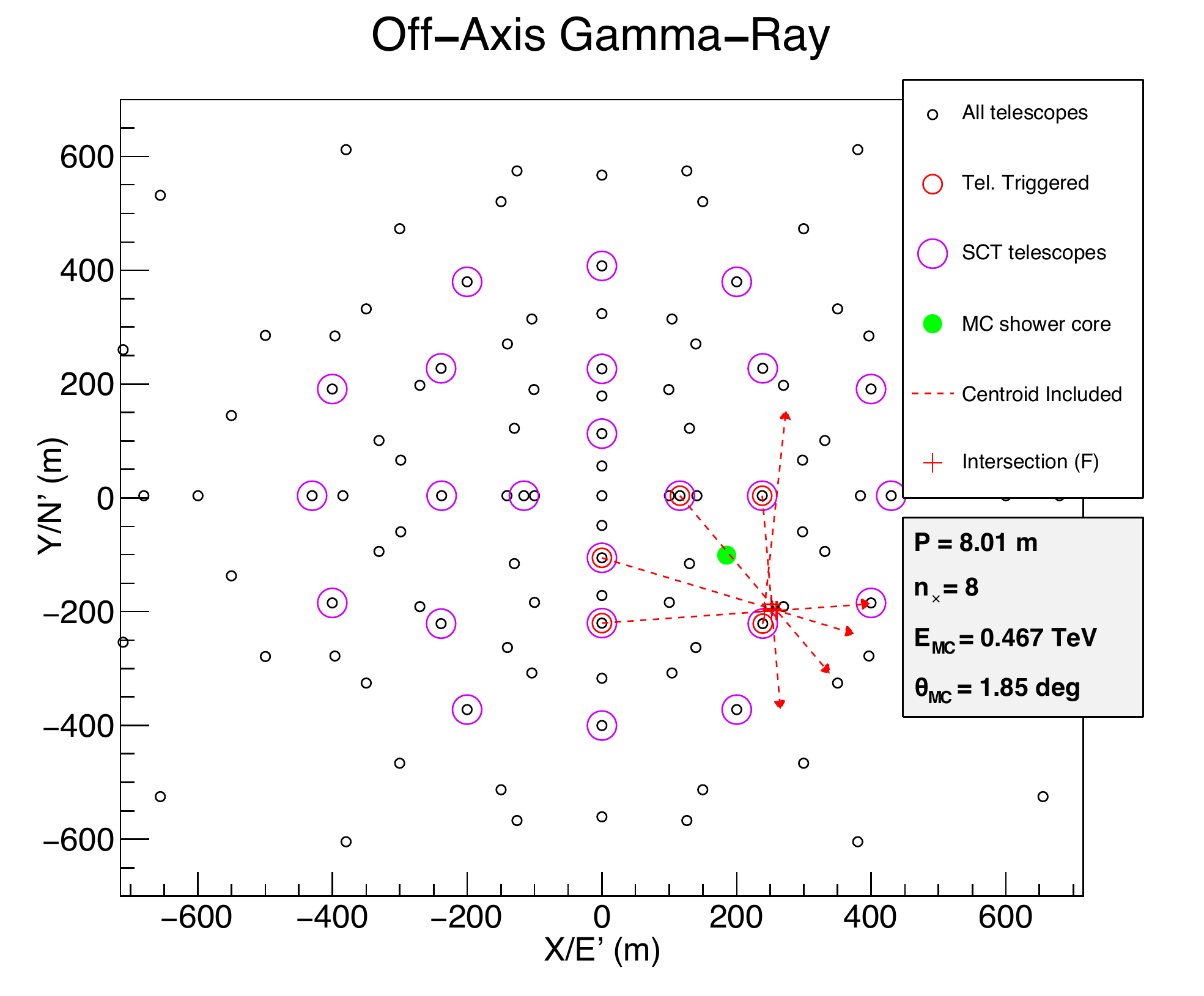}
 \caption{\small Array mirror plane projections illustrating $P$ computation for an on-axis \gr\ event (\textit{top}), and an off-axis \gr\ event with $E_{\gamma}\lesssim2$ TeV (\textit{bottom}). \textit{Red circles} identify SCT telescopes (\textit{magenta circles}) that triggered. \textit{Red dashed} arrows indicate the projected direction of $\mathbf{r}_{F,i}^{\prime}$ for each telescope in the array that generated a valid trigger-image. \textit{Red crosses} are used to indicate the set of valid \textit{forward} intersections $\{\mathbf{r}_{\times,j}^{\prime}\}$ between the $\{\mathbf{r}_{F,i}^{\prime}\}$. The green marker indicates the coordinates at which the shower core intersects the mirror plane. \textit{Grey panels} list the computed values of $P$, the number of {\textit{forward}} intersections {for which $20^{\circ} < \theta_{\times} < 160^{\circ}$} ($n_{\times}$) that were used in the computation, the \textit{true} energy ($E_{\rm MC}$) of the simulated event, and the \textit{true} angular offset ($\theta_{MC}$) of the between the incident particle direction and the array pointing.}\label{fig:P_computation_MP1}
\end{figure*}

\begin{figure*}[p]
 \center
 \includegraphics[width=0.68\textwidth]{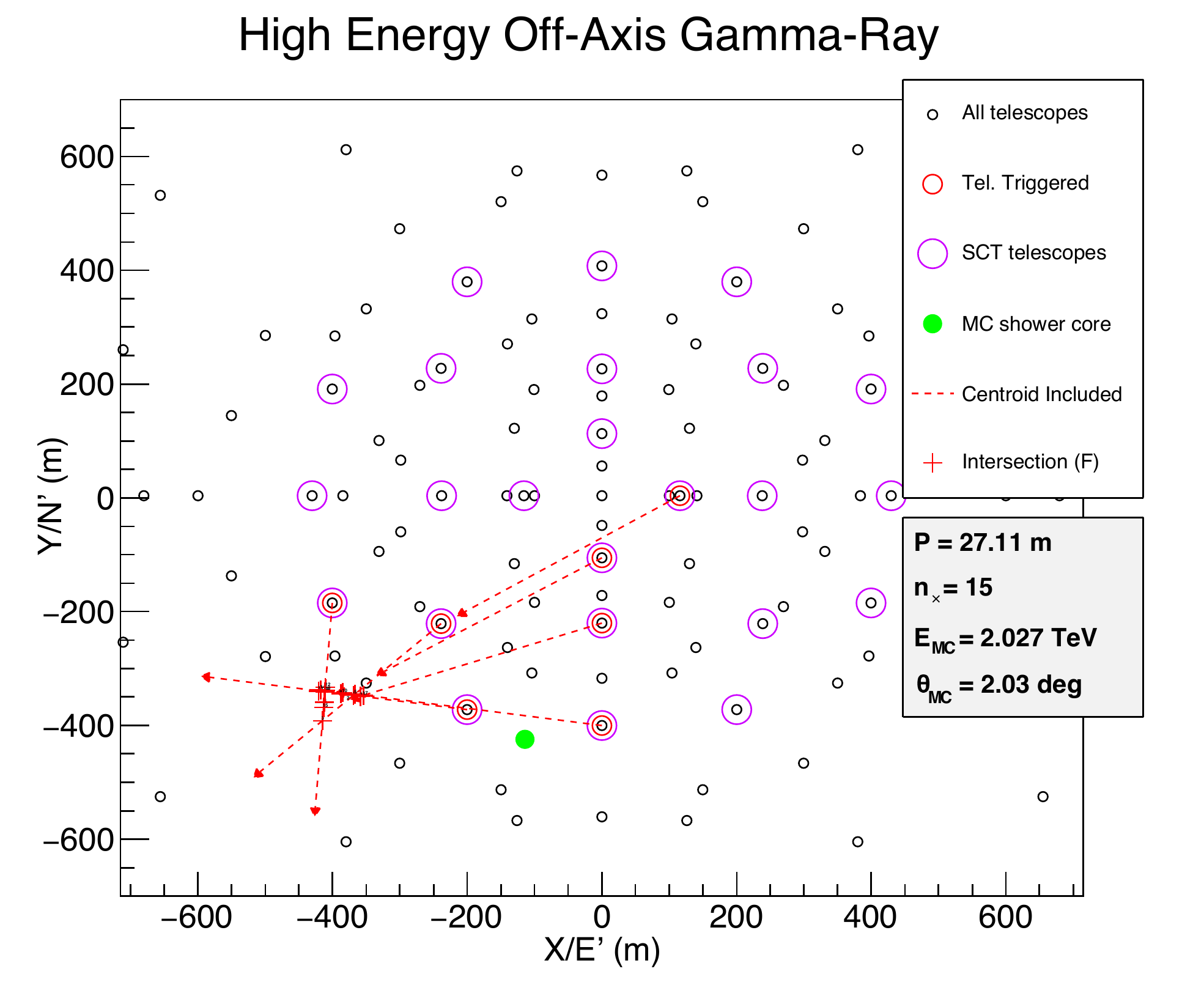}\\
 \includegraphics[width=0.68\textwidth]{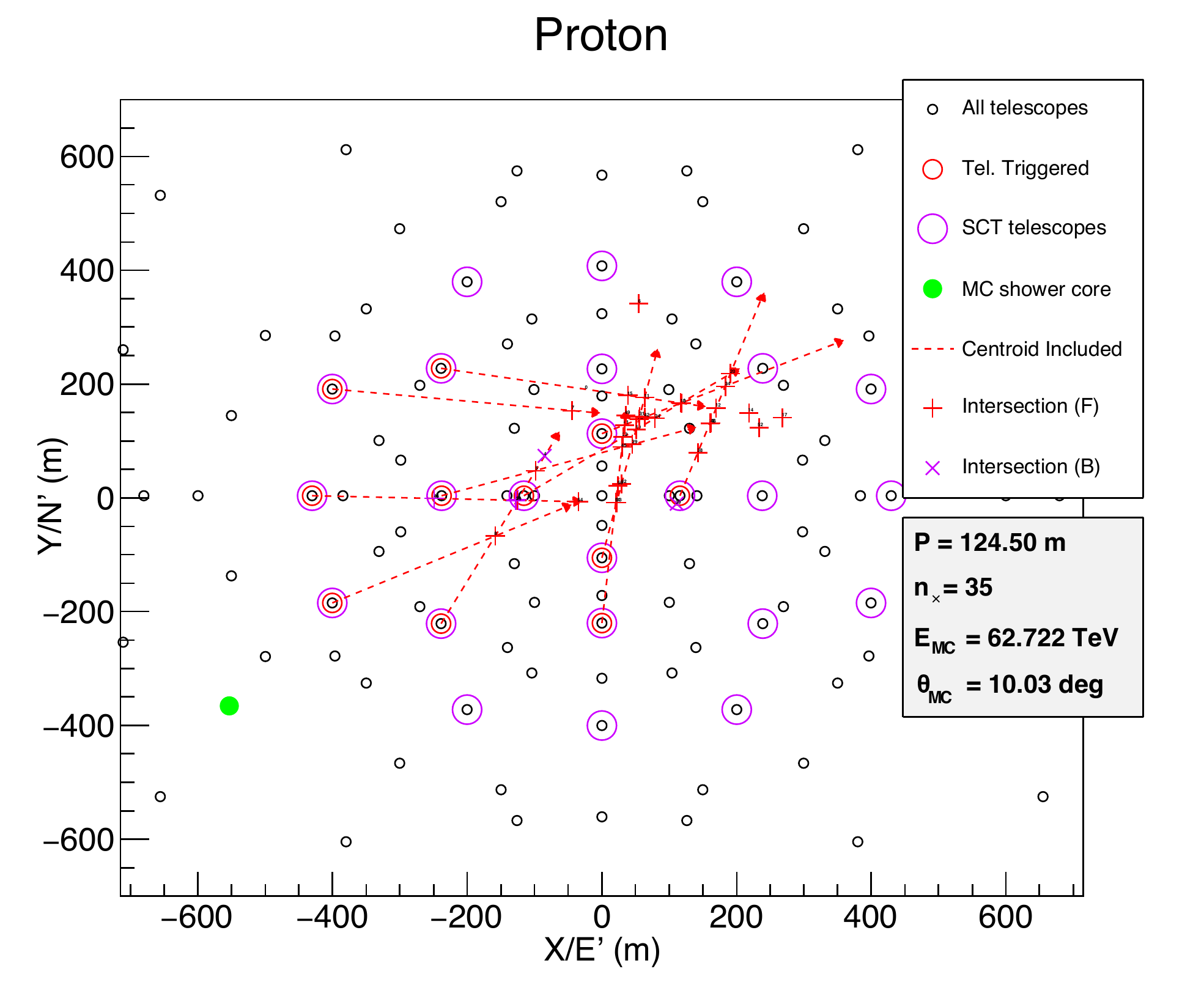}
 \caption{\small Array mirror plane projections illustrating $P$ computation for an off-axis \gr\ event for $E_{\gamma}\gtrsim2$ TeV (\textit{top}), and an off-axis cosmic-ray proton event (\textit{bottom}). See Figure \ref{fig:P_computation_MP1} for an explanation of the event-specific quantities that each marker is used to indicate.}\label{fig:P_computation_MP2}
\end{figure*}

Table \ref{tab:p_deriv_vars} summarizes the notation used in the subsequent derivation which also uses the index variable $i$ to enumerate the set of all triggered telescopes.
\begin{enumerate}
 \item For each telescope that triggers, the \textit{centroid}-vector $\mathbf{r}_{C,i}$ of the coarsely-sampled trigger image is defined as the mean camera-coordinates of the set of triggered super-pixels.
 \item For each centroid vector, a second vector $\mathbf{r}_{F,i} = \mathbf{r}_{C,i} - \mathbf{r}^{\star}$ is defined where $\mathbf{r}^{\star}$ is an arbitrarily selected \textit{fiducial} camera-plane coordinate $\mathbf{r}^{\star} = (x_{C}^{\star}, y_{C}^{\star})$.\footnote{The \textit{direction} of this vector is hereafter described as the \textit{forward} direction and intersections between pairs of vectors along their forward directions are described as \textit{forward intersections}.} For real telescope arrays, the introduction of $\mathbf{r}^{\star}$ provides the flexibility to correct for non-uniform mechanical deformations of individual telescope structures, or non-parallel telescope pointing modes, and ensure that $\mathbf{r}_{F,i}$ corresponds to the projection of identical celestial coordinates in all telescope cameras. This study simulates all telescopes identically and computation of $P$ is simplified by defining $\mathbf{r}^{\star}$ to be the camera centre $(0,0)$.
\end{enumerate}
The events shown in each panel of Figures \ref{fig:P_computation_MP1} and \ref{fig:P_computation_MP2} provide representative examples of \gr\ and proton-initiated events with various simulated characteristics. They illustrate how the subsequent algorithmic operations are used to compute $P$ for each event class.
\begin{enumerate}
 \item The direction of each computed $\mathbf{r}^{\star}_{F}$ vector is projected from the independent coordinate systems of each telescope camera plane into a unified coordinate system spanning the \textit{mirror plane} of the telescope array, defining a set of \textit{projected} vectors $\{\mathbf{r}_{F,i}^{\prime}\}$. The mirror plane is defined to intersect the mean of the geographical coordinates of all telescopes comprising the complete array and to lie perpendicular to their common pointing axes.\footnote{{Accordingly, \textit{if} the telescopes of the array were coaligned to point vertically upwards, then the mirror plane and ground plane would be identical.}}
 \item A set of $n_{\times}$ mirror-plane coordinates $\{\mathbf{r}_{\times,j}^{\prime}\}$ is computed that corresponds to the \textit{forward} intersections between the projected $\{\mathbf{r}_{F,i}^{\prime}\}$ vectors. {Only intersections that subtend angles $20^{\circ} < \theta_{\times} < 160^{\circ}$ are used for the computation of $P$\footnote{Intersections that subtend angles outside of this range are excluded since discretization of the camera image induces small errors affecting the computation of the $\{\mathbf{r}_{F,i}^{\prime}\}$ which can amplify to produce large, spurious offsets when the $\{\mathbf{r}_{\times,j}^{\prime}\}$ are computed.}.} The index variable $j$ is used to enumerate the set of all forward intersections.
 \item Finally, the \textit{parallax width} is defined as the dispersion of the intersection coordinates
       \begin{equation}\label{eq:parwidth_expression}
        P = \sqrt{\frac{\sum_{j}\left| \mathbf{r}_{\times,j}^{\prime} - \langle \mathbf{r}_{\times,j}^{\prime} \rangle \right|^{2}}{n_{\times}}}.
       \end{equation}
\end{enumerate}
The \textit{upper}, \textit{lower} panels of Figure \ref{fig:P_computation_MP1} and the \textit{upper} panel of Figure \ref{fig:P_computation_MP2} all correspond to \gr-initiated events. For such events, the computed trigger-image centroids $\mathbf{r}_{C,i}$ correspond closely with the camera-plane projection of a single point in 3-dimensional space. This point lies close to the air-shower axis at the height of maximal Cherenkov emission. The fiducial camera coordinate $\mathbf{r}^{\star} = (0,0)$ is also the camera-plane projection of a second, infinitely distant point that is identical for all telescopes. For \gr-initiated events,  each $\mathbf{r}_{F,i}$ connects projected points that are effectively identical for all telescopes. Accordingly, the mapping $\{\mathbf{r}_{F,i}\}\rightarrow\{\mathbf{r}_{F,i}^{\prime}\}$ yields a set of vectors which intersect at a tightly clustered region of the array mirror plane and the computed value of $P$ is small.

Figure \ref{fig:P_computation_MP1} (\textit{top}) corresponds to a \gr-initiated event aligned with the telescopes' optical axes (\textit{on-axis}).  The arrival directions of on-axis events correspond precisely with the fiducial camera-plane coordinate $\mathbf{r}^{\star} = (0,0)$, when projected into the camera plane's coordinate system\footnote{All points along the event's arrival direction-vector are implicitly \textit{on the air-shower axis} and that $\mathbf{r}^{\star}$ corresponds to the projection of the point on that vector that is located at infinity}. Consequently, the derived $\{\mathbf{r}_{\times,j}^{\prime}\}$ coincide closely with the projection of the axis into the array mirror plane.
The \textit{lower} panel of Figure \ref{fig:P_computation_MP1} is representative of an \textit{off-axis} \gr\ event with energy \textit{below} 2 TeV. Low energy events typically produce compact camera images with small offsets between the true image centroid and that of the Boolean-valued trigger image. Accordingly, the projected point pairs connected by each of the $\mathbf{r}^{\star}_{F}$ remain \textit{almost} identical for all telescopes and the tight clustering of $\{\mathbf{r}_{\times,j}^{\prime}\}$ is preserved. The arbitrary, \textit{a-priori} definition of $\mathbf{r}^{\star}$ implies that the single, infinitely distant point that is projected to those camera-coordinates for \textit{all} telescopes will \textit{generally not} lie on the air shower axis. Accordingly, the coincidence between the  $\{\mathbf{r}_{\times,j}^{\prime}\}$ and the mirror-plane projection of the air-shower axis is lost. A more detailed discussion of this effect, as well as some additional considerations that apply to higher energy off-axis gamma-ray events, is presented in $\S$\ref{sec:passthrough}.

Finally, the \textit{lower} panel of Figure \ref{fig:P_computation_MP2} illustrates a proton-initiated event. There is now no guarantee that $\mathbf{r}_{C,i}$ for each telescope corresponds with the projection of a single 3-dimensional point, since different telescopes may image multiple, different sub-showers of the hadronic cascade. Accordingly, the tight clustering of the $\{\mathbf{r}_{\times,j}^{\prime}\}$ is lost and the computed value of $P$ is large.

\section{Indiscriminate Acceptance for High-Energy Events}\label{sec:passthrough}
The parallax width algorithm assumes that the centroids of \textit{trigger} images reliably encode the geometry of the air-showers.
If the intensity profile of the Cherenkov light image is substantially as{s}ymetric, the validity of this assumption may degrade. Ideally, the trigger-image centroid $\mathbf{r}_{C,i}$ should correspond closely with the \textit{full image centroid} $\mathbf{r}_{C^{\dag},i}$, defined as the signal-amplitude-weighted mean position of all \textit{imaging} pixels that trigger in response to incident Cherenkov photons. Without access to the information provided by the individual pixel \textit{amplitudes}, the Boolean-valued \textit{trigger} images appear more symmetric and the $\mathbf{r}_{C,i}$ that are used to compute $P$ may not accurately represent the air-shower geometry.

The \textit{left} and \textit{right-hand} panels of Figure \ref{fig:P_computation_Camera} illustrate schematically how any camera-coordinate offsets $\epsilon_{C,i} = \mathbf{r}_{C,i} - \mathbf{r}_{C^{\dag},i}$ between the two centroid definitions produce corresponding directional perturbations of each $\mathbf{r}_{F,i}$. A representative mirror-plane projection of these misaligned vectors is illustrated in the \textit{top panel} of Figure \ref{fig:P_computation_MP2}, and yields a set $\{\mathbf{r}_{F,i}^{\prime}\}$ that typically increases the dispersion between the $\{\mathbf{r}_{\times,j}^{\prime}\}$ intersection coordinates, and inflates the computed value of $P$. The potential magnitude of $\epsilon_{C,i}$ increases for high-energy \gr-initiated air showers, which typically produce extensive images that comprise a large number ($n_{\rm TP}$) of triggered super-pixels.

\begin{figure*}[p]
 \centering
 \includegraphics[width=0.7\textwidth]{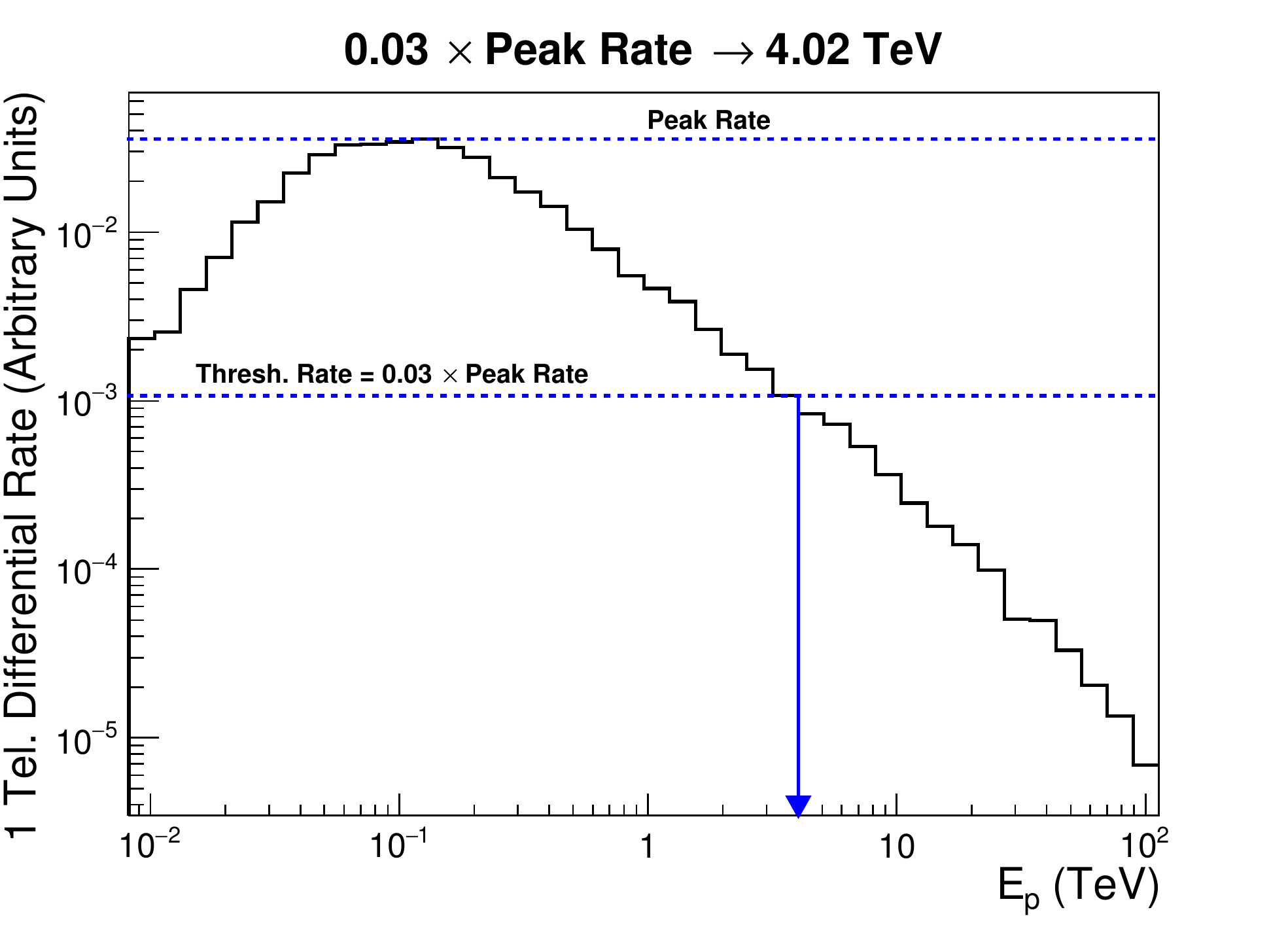}\\
 \includegraphics[width=0.7\textwidth]{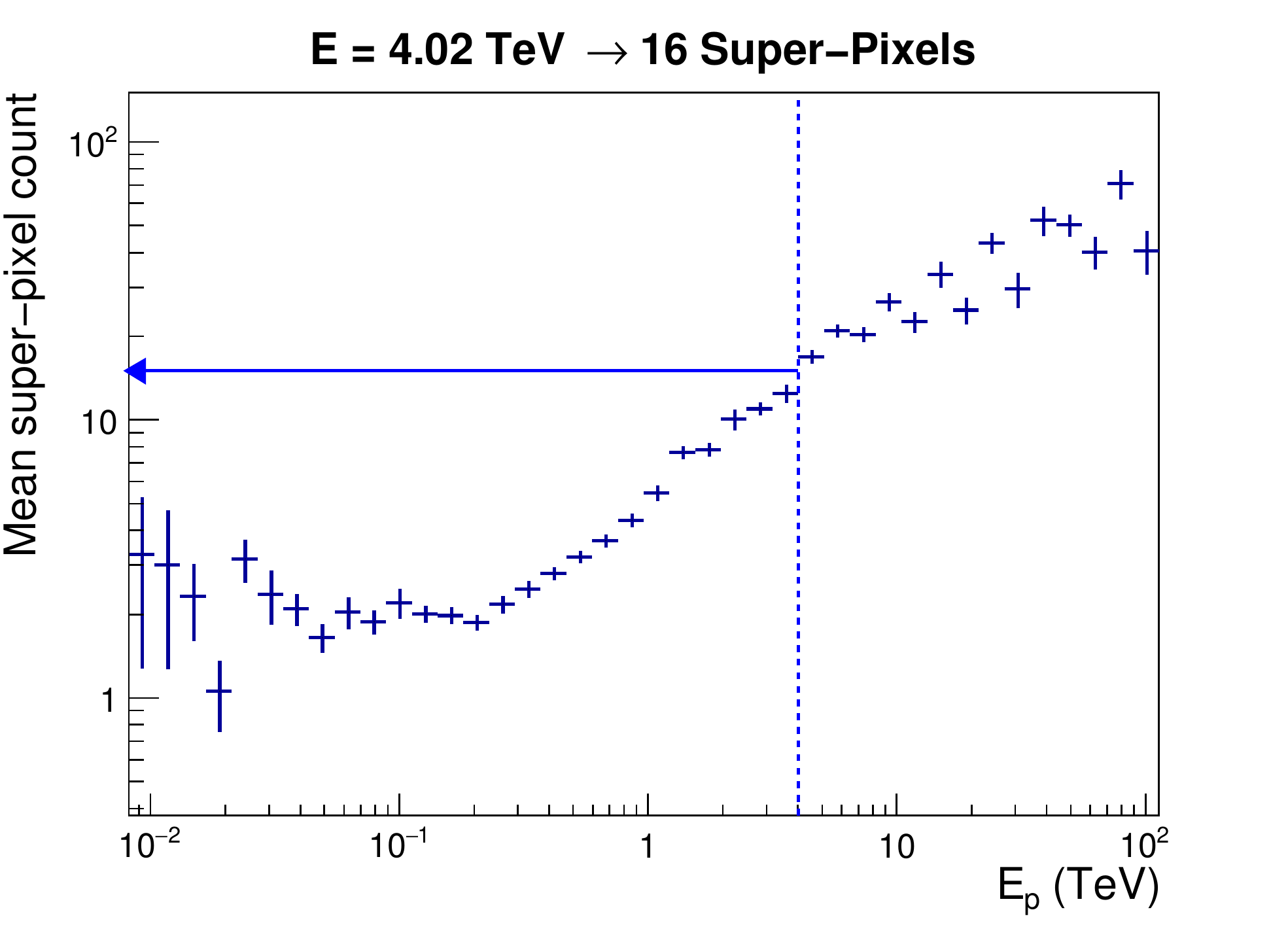}
 \caption{\small Computation of the super-pixel multiplicity $n^{\rm pass}_{\rm TP}$ that is required to satisfy the pass-through criterion. In the \textit{top} panel, the energy-binned effective collection area for \textit{proton}-initiated events that trigger \textit{at least one} telescope is folded with the expected cosmic ray spectral shape ($\propto E_{p}^{-2.7}$) and used to determine the typical proton energy $E_{p,3\%}$ at which the expected rate of \textit{single} telescope triggers falls below 3\% of its peak value. Note that \gr-like images of a hadronic sub-shower typically sample a fraction of the energy of the incident proton. Accordingly, genuine \grs\ that produce images comprising $n^{\rm pass}_{\rm TP}$ trigger pixels have energies that are typically $\sim30\%$ that of protons that do so. In the \textit{bottom} panel, the energy-binned distribution of mean trigger image super-pixel multiplicities is used to derive the value of $n^{\rm pass}_{\rm TP}$ that corresponds to events for which $E_{p}\sim E_{p,3\%}$.}\label{fig:passthrough_comp}
\end{figure*}

To prevent spurious rejection of \textit{genuine} high-energy, \gr-initiated events, a pre-calibrated multiplicity threshold $n_{\rm TP} = n^{\rm pass}_{\rm TP}$ is used to unconditionally accept (or \textit{pass through}) events for which \textbf{any} telescope in the array captures a trigger image comprising $n^{\rm pass}_{\rm TP}$ or more super-pixels. As illustrated by Figure \ref{fig:passthrough_comp}, the expected \textit{single-telescope} trigger rate $R_{p}$ for simulated, \textit{proton}-initiated air-showers is used to calibrate an appropriate value for $n^{\rm pass}_{\rm TP}$. To retain effective suppression of the most frequent cosmic-ray triggers, a threshold corresponding to the typical super-pixel multiplicity for \textit{proton}-initiated events that trigger at 3\% of the peak single-telescope rate is adopted. \revision{}{Data will be serialized and readout will occur for any telescope that records an image with $n_{\mathrm{TP}} > n_{\mathrm{TP}}^{\mathrm{pass}}$. Accordingly, in order to select a threshold that effectively controls the single telescope dead time induced by cosmic-ray events that fulfill the pass-through criterion, it is important to calibrate $n_{\mathrm{TP}}^{\mathrm{pass}}$ using the differential trigger rate for a \textit{single} telescope.}

\section{Hardware Implementation of the Parallax Width Algorithm}\label{sec:hw_implementation}

The simplicity of the algorithm that calculates $P$ was intentionally designed to facilitate rapid, FPGA-based computation \citep[See e.g.][for examples of existing VHE \gr\ telescopes that use similar technology]{4774947, 2013arXiv1307.8360Z}. {Rapid exchange of locally synthesized trigger-images and high-resolution timing data between nearby telescopes} can be used to derive independent, telescope-specific values of $P$ that can inform the decision to initiate or veto readout of finely pixelated image data. The ability to intelligently filter events before they are fully digitized, transmitted and stored \revision{also eliminates any}{substantially reduces the} system deadtime that would be associated with those processes. \revision{}{We note that the parallax width veto does not completely eliminate system dead time, since events that are \textit{accepted} by the array trigger will still require digitization and readout.}

The design of the SCT prototype includes a $16\,\mu{\rm s}$ pixel readout buffer \citep{2015arXiv150902345O,2015arXiv150806296T}, which exceeds the combined time required to exchange data between telescopes and compute $P$, enabling high-resolution signal data to be temporarily stored by each telescope pending a trigger decision. Hardware and firmware that implement the DIAT triggering scheme have also been developed and will be deployed concurrently with the SCT prototype. {Hardware-based computation of $P$ by direct evaluation of \eqref{eq:parwidth_expression} is not practical. A proven alternative approach entails straightforward combination of values that are extracted from precomputed lookup tables \citep[e.g.][]{4774947,2013arXiv1307.8360Z}. Moreover, provided with appropriately parameterised tables, factors that complicate the computation, such as telescope field rotation or structural deformation under slewing, can be effectively addressed.}

\section{Results and Discussion}\label{sec:results}

\begin{figure*}
 \centering
 \includegraphics[width=0.7\textwidth]{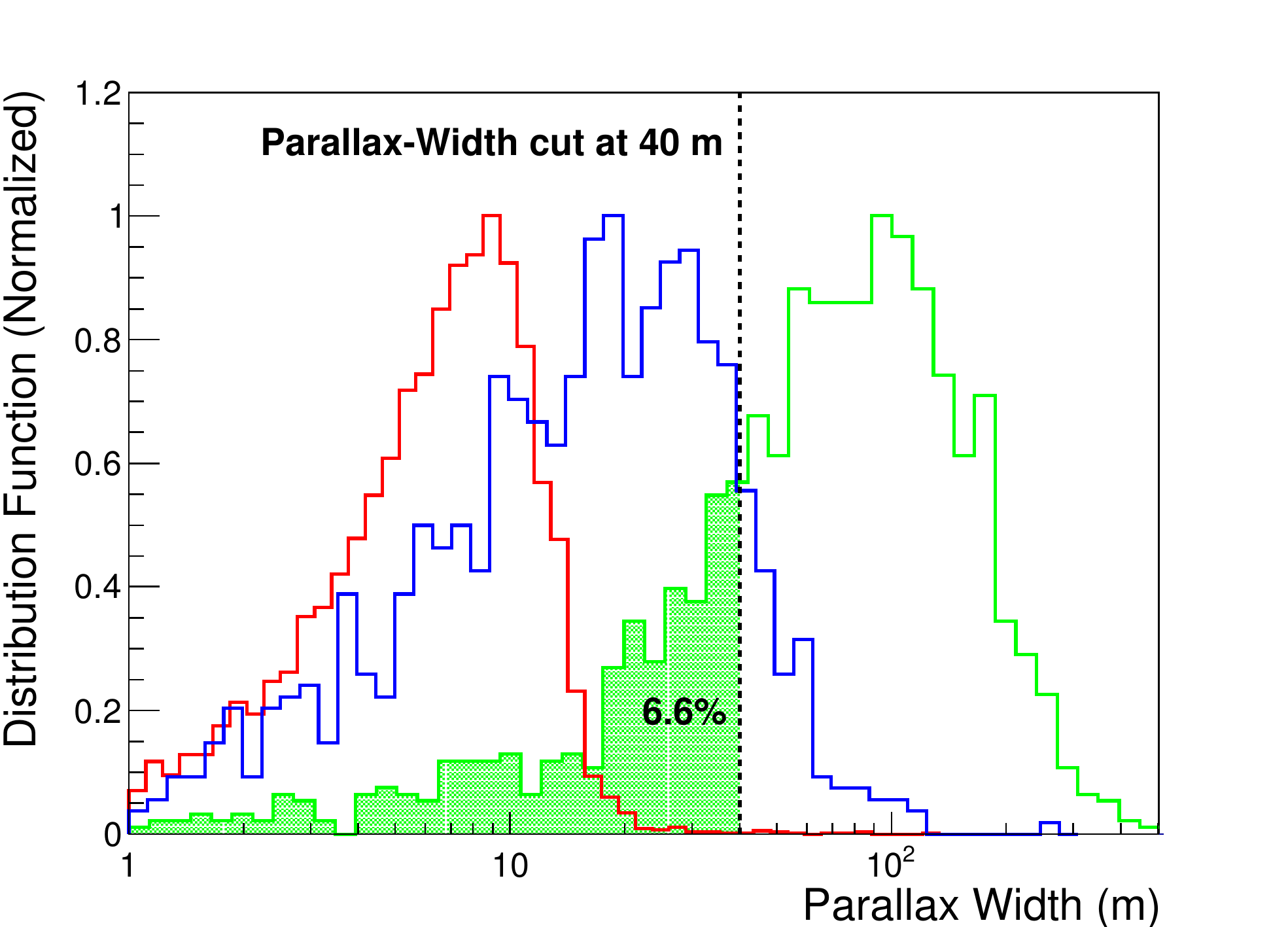}\\
 \includegraphics[width=0.7\textwidth]{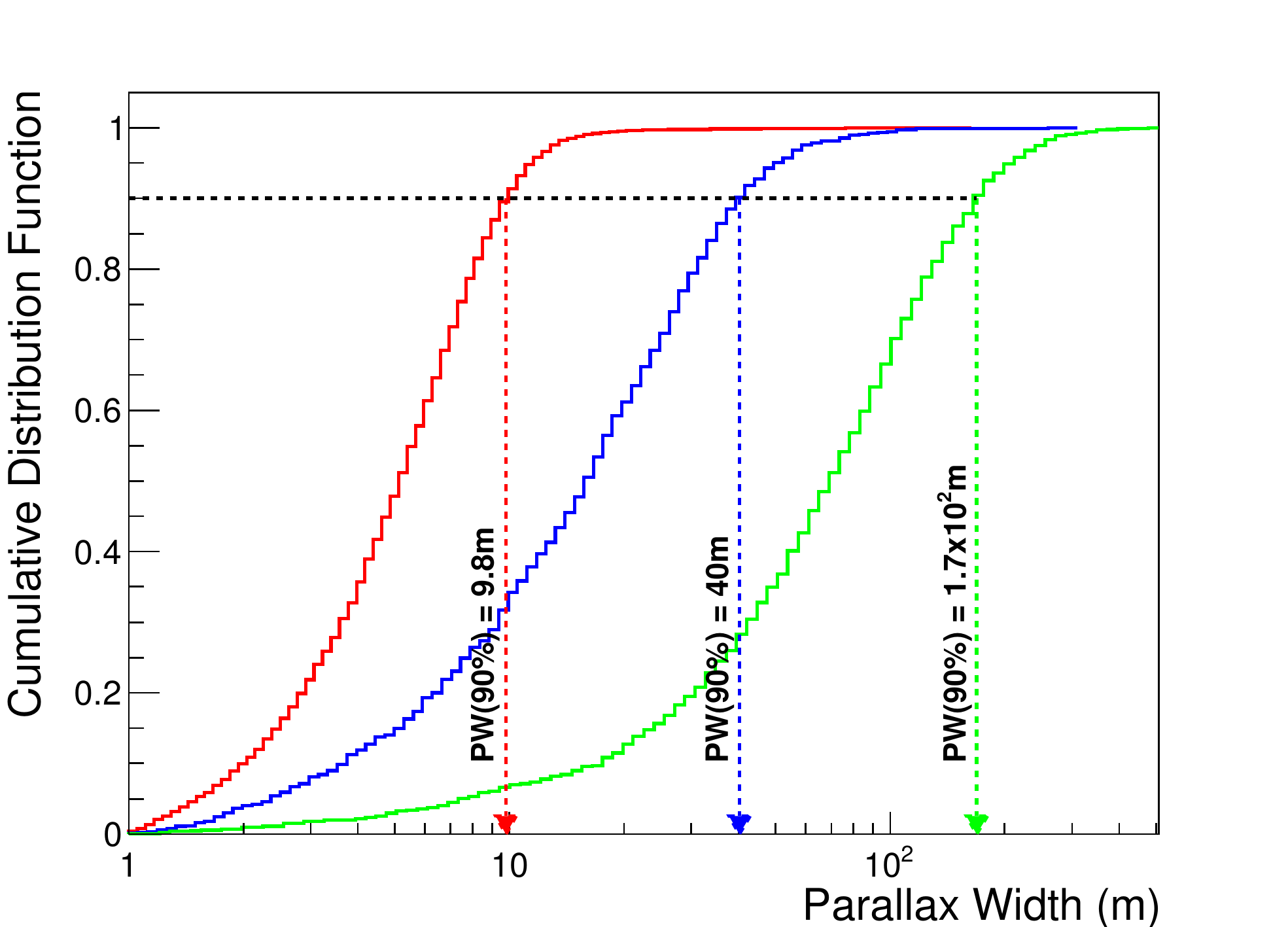}
 \caption{\small The computed distributions (\textit{top}) and CDFs (\textit{bottom}) of $P$ values corresponding to point-origin  (on-axis, \textit{red}) and diffuse (\textit{blue}) \gr\ events, and diffuse cosmic-ray background events (\textit{green}).}\label{fig:pwidth_dists}
\end{figure*}

Figure \ref{fig:pwidth_dists} (\textit{top}) shows the three distributions of $P$ values that are computed for each investigated dataset, while the \textit{bottom} panel displays the corresponding cumulative distributions for $P$. For the point-like \gr\ source, all the simulated photons are incident on-axis, and 90\% of computed $P$ values are  $<11$ m, in accordance with expectation. For extended \revision{}{and diffuse} \gr\ sources, the majority of incident photons are incident off-axis. Nonetheless, the expected prevalence of small $P$ values is realized, with 90\% of events that do \textit{not} fulfil the pass-through criterion having $P<40\,{\rm m}$. In contrast $\sim 93\%$ of all proton-initiated events yield $P>40\,{\rm m}$, confirming the expectation that $P$ is a highly efficient discriminator between \gr- and proton-initiated events.

\begin{figure*}
 \centering
 \includegraphics[width=0.7\textwidth]{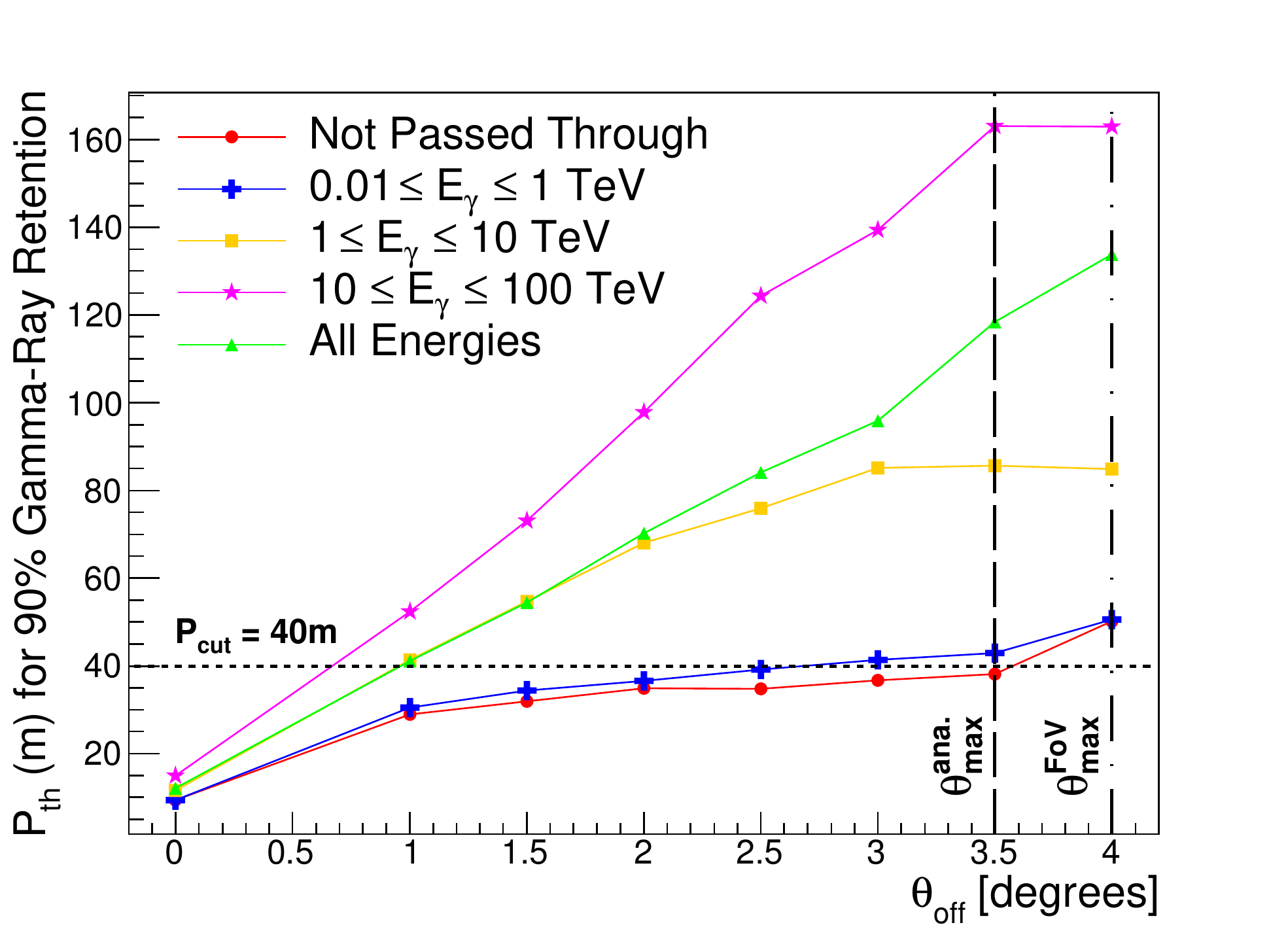}
 \caption{\small Evolution of \gr\ retention with increasing angular offset $\theta_{\rm off}$ between the common optical axes of all telescopes in the array and true air-shower axis. The various curves represent different subsets of the simulated event ensemble and illustrate the threshold value $P_{\rm th}(\theta_{\rm off})$  \textit{below} which 90\% of \grs\ that trigger \textit{at least three} telescopes in the array are retained by the selection algorithm. The \textit{red} curve corresponds to predominantly low-energy \gr\ events that would \textbf{not} fulfill the pass-through array trigger criterion. The \textit{blue}, \textit{yellow} and \textit{magenta} curves represent energy-selected subsets and illustrate that \gr\ retention at large offsets degrades with increasing \gr\ energy ($E_{\gamma}$). Finally, the \textit{green} curve represents the union of the energy-selected subsets.  Reliable reconstruction of events that are incident at offsets in excess of $\theta_{\max}^{\rm ana.} = 3.5^{\circ}$ (\textit{dashed vertical} line) is typically problematic and such events would be discarded by subsequent data analysis. The \textit{dot-dashed vertical} indicates the angular extent ($\theta_{\max}^{\rm FoV} = 4^{\circ}$) of the SCT camera.}\label{fig:pw_thresh_vs_offset}
\end{figure*}

Figure \ref{fig:pw_thresh_vs_offset} illustrates the evolution of \gr\ event retention with increasing angular offset $\theta_{\rm off}$ between the assumed \gr\ incidence direction and the telescope optical axis. A point-like \gr\ source was simulated at various offsets from the camera centre and the resultant data were used to derive the threshold parallax width value of $P_{\rm th}$ that \textit{retains} 90\% of events  which yield three valid trigger images. The expected energy dependence of $P_{\rm th}$ (see $\S$\ref{sec:passthrough} for details) is evident, with event retention degrading most rapidly at large offsets for \gr\ energies exceeding 10 TeV but remaining $\lesssim40\,{\rm m}$ for $E_{\gamma} < 1$ TeV. {Critically, the lowest energy event subset is characterized by the faintest Cherenkov images and therefore overlaps significantly with the set of events for which \textit{none} of the triggering telescopes fulfill the pass-through criterion ($n^{\rm pass}_{\rm TP} = 16$).} {An important corrolary is that those higher energy, large-offset events that \textit{would} have been rejected on the basis of $P$ alone are retained if the pass-through criterion is re\revision{}{s}pected.}

{For the simulated array, indiscriminately accepting all events that fulfill the pass-through criterion, and adopting a $P_{\rm th} = 40\,{\rm m}$ for the remainder results in retention of $\gtrsim 90\%$ of genuine \gr\ events for $\theta_{\rm off} \lesssim 3.5^{\circ}$.}

\begin{figure*}
 \centering
 \includegraphics[width=0.7\textwidth]{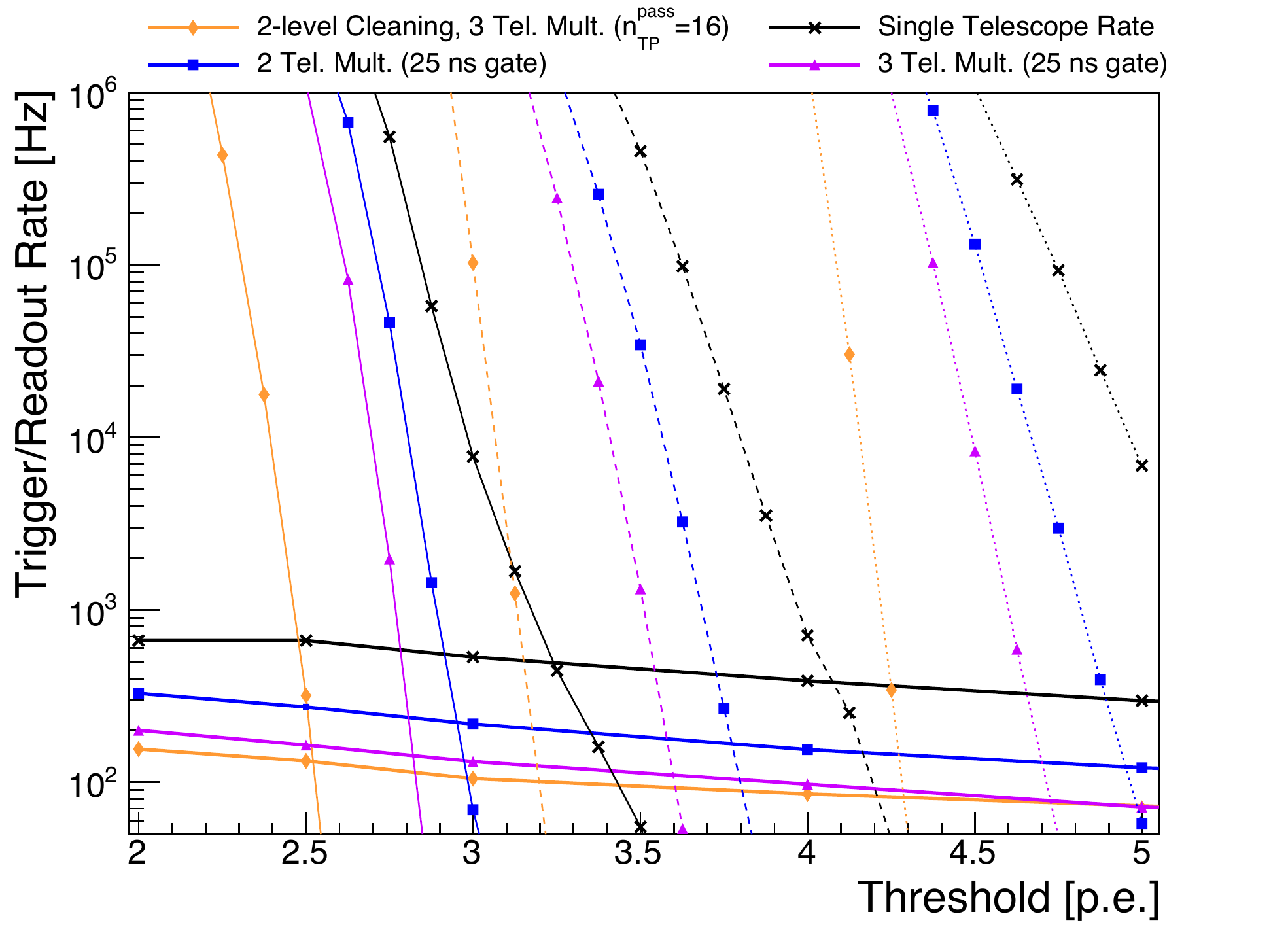}
 \caption{\small Projected \textit{single telescope} readout rates following  cosmic-ray or NSB-induced array triggers for various simulated super-pixel trigger thresholds. The expected array trigger rate in the absence of a multi-telescope trigger criterion is illustrated by the \textit{black crosses}. The \textit{blue squares} are derived by requiring that \textit{at least} two telescopes in the array trigger within $25\,{\rm ns}$, while \textit{magenta triangles} illustrate the effect of increasing the telescope multiplicity requirement from two to three. \textit{Orange diamonds} illustrate rates that result when the requirement that \textit{at least} three telescopes trigger within $25\,{\rm ns}$ is retained before application of the two-level cleaning algorithm and subsequent indiscriminate acceptance of events with $n_{\rm TP} \geq 16$. 
  \textit{Thick, solid curves} indicate the expected readout rate of cosmic-ray\revision{}{-proton}-induced triggers. The remaining curves illustrate the expected readout rate of NSB-induced triggers for nominal (\textit{thin solid lines}), two-times-nominal (\textit{dashed lines}) and four-times-nominal (\textit{dotted lines}) NSB intensity levels. The nominal NSB level \revision{}{implies production of 11.96 photoelectrons per super-pixel, per microsecond, which} is consistent with a typical extragalactic field of view.}\label{fig:crAndNsbBiasCurves}
\end{figure*}

Figure \ref{fig:crAndNsbBiasCurves} illustrates how the rate of background-induced telescope triggers increases as the super-pixel trigger threshold is reduced. The rates were derived using simulations of proton-initiated air showers and random incidence of NSB photons upon each camera pixel is modelled by the detector simulation software. Individual curves represent the frequency of single-telescope triggers that are subsequently retained by a particular array triggering strategy. The requirement for retention by an array trigger implies a direct correspondence between the derived rates and the average frequency with which readout of finely pixelated image data would be initiated for each telescope.

The upper four curves in Figure \ref{fig:crAndNsbBiasCurves} share a number of distinctive characteristics. For small $n_{\rm pe}$, the majority of single telescope triggers are induced by random triggering of camera pixels by NSB photons. The probability that random pixel triggers will generate a valid telescope trigger decreases rapidly with increasing $n_{\rm pe}$ and the resultant rate curve is characterized by a steeply falling power law. An \textit{inflection point} marked by a reduction of the absolute power law index at a particular threshold ($n^{\star}_{\rm pe}$) represents the transition to a regime in which Cherenkov photons from cosmic-ray-induced air showers dominate the single telescope trigger rates. Whereas the incident flux of cosmic rays is typically stable and isotropic for a particular telescope location, the NSB intensity can vary markedly between different astronomical fields, and sporadic ambient light sources may also illuminate telescopes without warning. Figure \ref{fig:crAndNsbBiasCurves} illustrates that \revision{}{without an array trigger, reliably stable operation would require $n^{\star,\mathrm{1-fold}}_{\rm pe} \gtrsim 3.5\,\mathrm{ p.e.}$. Relative to this threshold, \textit{any} array trigger schemes enable reduction of $n^{\star}_{\rm pe}$ and thereby reduce the threshold energy of \grs\ that the instrument can detect. Moreover, while traditional 2- and 3-fold multiplicity array triggers yield $n^{\star,\mathrm{2-fold}}_{\rm pe}\approx{3.0}\,\mathrm{p.e.}$ and $n^{\star,\mathrm{3-fold}}_{\rm pe}\approx{2.7}\,\mathrm{p.e.}$}, a simulated array trigger that implements the parallax width algorithm and unconditionally accepts events that fulfill $n_{\rm TP} \geq 16$, lowers the threshold at which the inflection point occurs to $n^{\star}_{\rm pe}\approx{2.5}$$\,{\rm p.e}$\@. Specifying a pixel threshold \revision{$n_{\rm pe}\gg n^{\star}_{\rm pe}$}{$n_{\rm pe}$ that exceeds $n^{\star}_{\rm pe}$ by factors of at least a few} enables stable array operation in all but the most extreme observational conditions, but sacrifices sensitivity to intrinsically faint, low-energy \gr\ events. By substantially reducing $n^{\star}_{\rm pe}$, the parallax width trigger algorithm also lowers the overall energy threshold of the instrument.

\begin{figure*}[p]
 \centering
 \includegraphics[width=0.7\textwidth]{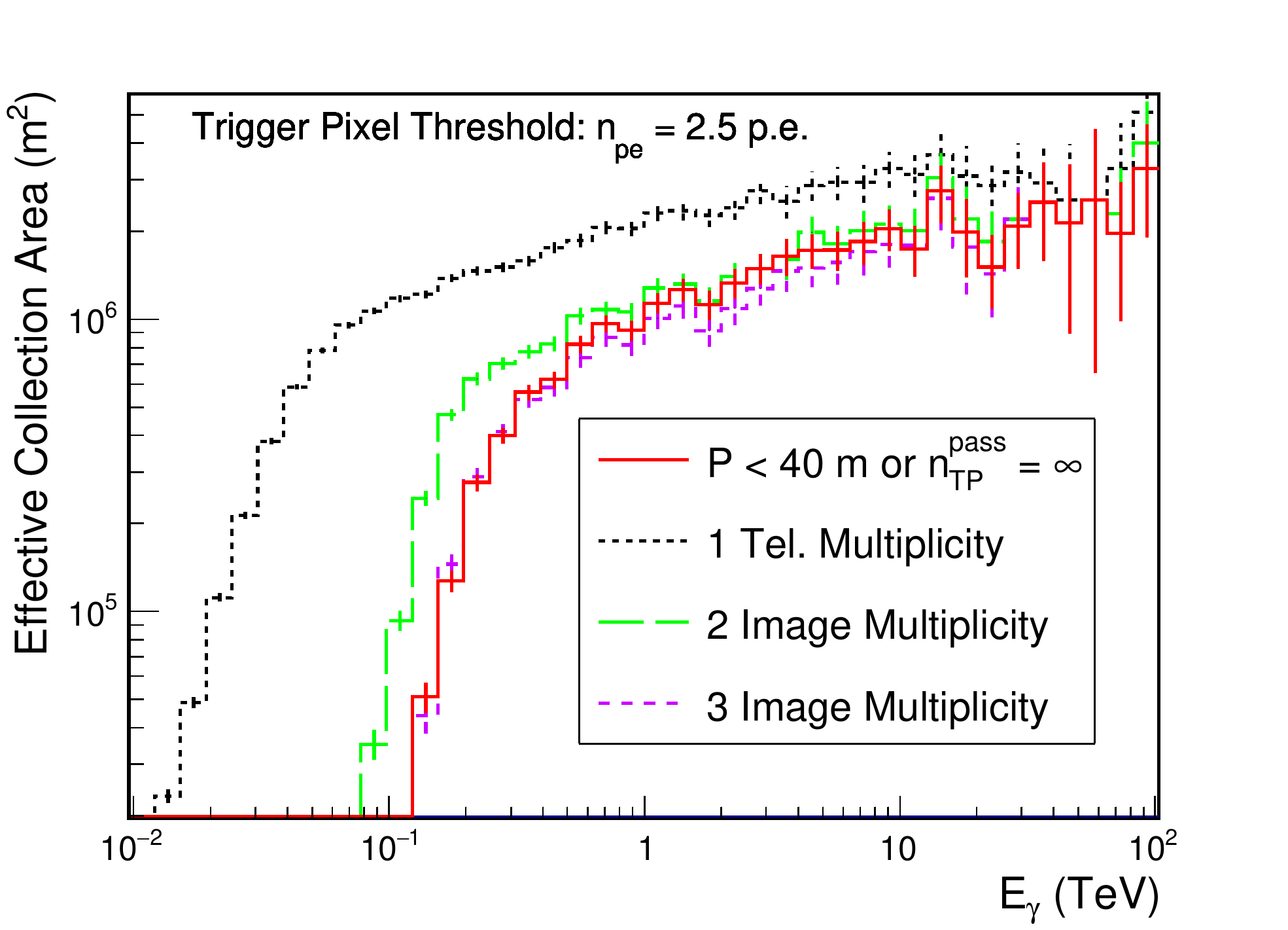}\\
 \includegraphics[width=0.7\textwidth]{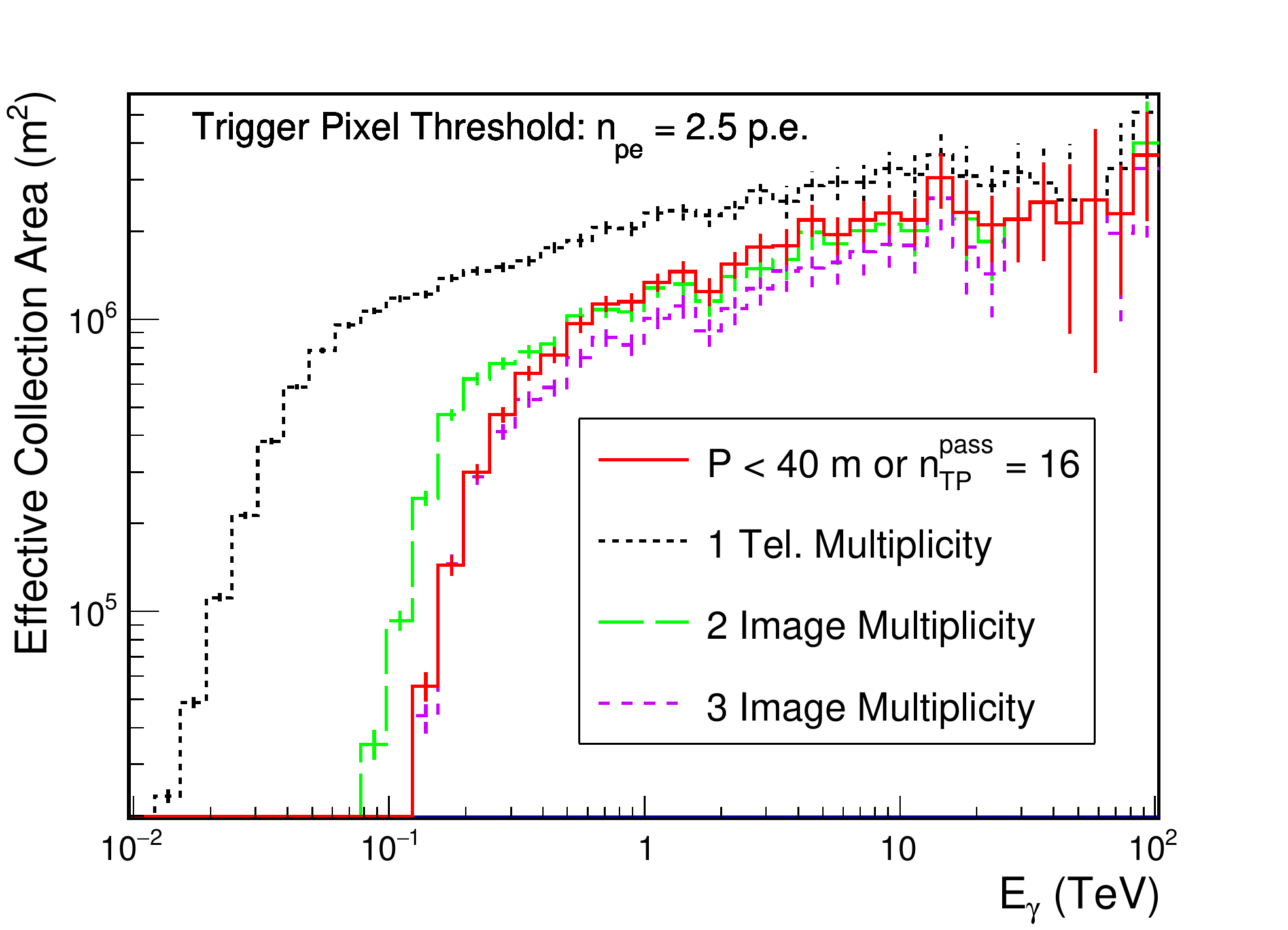}
 \caption{\small Collection areas for a point-origin \gr\ source and a super-pixel trigger threshold of {2.5}$\,{\rm p.e.}$ assuming $n^{\rm pass}_{\rm TP}\rightarrow\infty$ (\textit{top}), and  $n^{\rm pass}_{\rm TP}=16$ (\textit{bottom}). The \textit{black} curve represents a single telescope trigger requirement (i.e. no array trigger). The \textit{green} curve plots the collection area for events that trigger at least two telescopes and are subsequently found to yield two Cherenkov images that are useable for parameterization and geometric reconstruction of the corresponding air shower and the properties of its progenitor \gr.
  The \textit{violet} curve illustrates the collection area for a more desirable subset of events that yield \textit{at least three} high-quality Cherenkov images. The \textit{red} curve represents the  collection area \revision{}{for events that are accepted} after application of the parallax width criterion, rejecting any events for which $P>40\,{\rm m}$.}\label{fig:gamma_aeffs}
\end{figure*}

\begin{figure*}[p]
 \centering
 \includegraphics[width=0.7\textwidth]{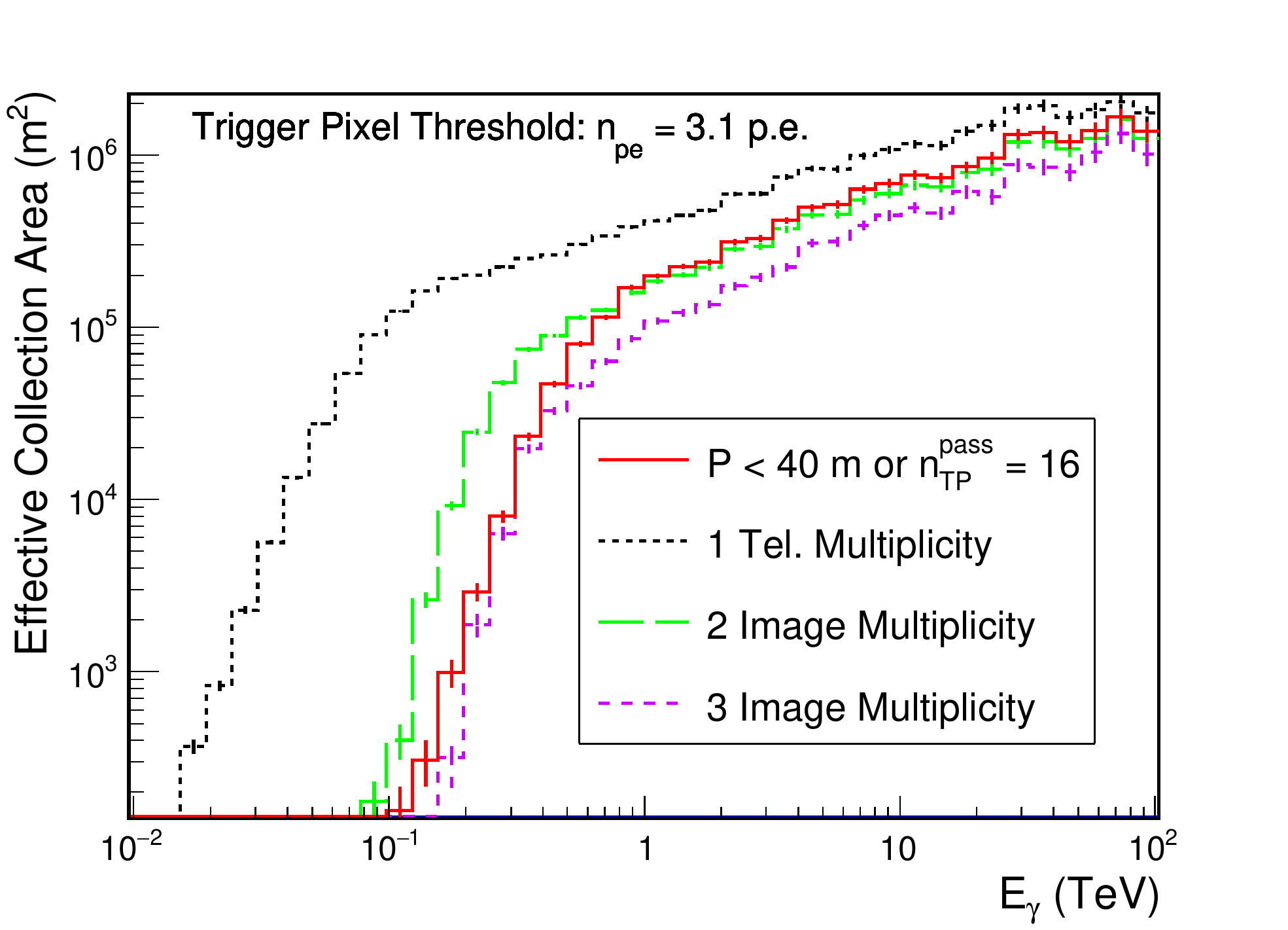}\\
 \includegraphics[width=0.7\textwidth]{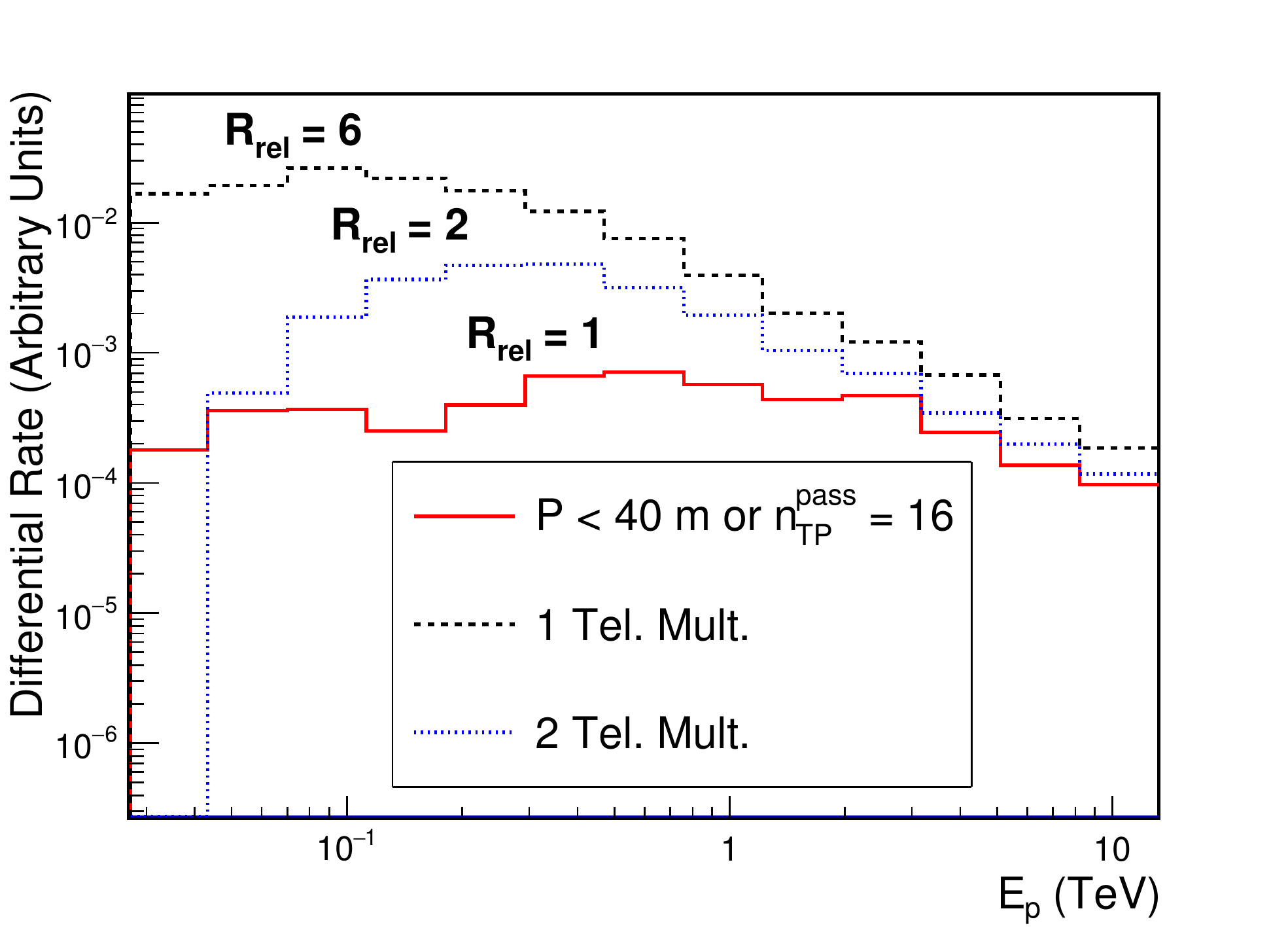}
 \caption{\small \textit{Top panel:} Collection areas for diffuse \gr\ emission that correspond to different array triggering strategies\revision{}{, assuming a super-pixel trigger threshold of {2.5}$\,{\rm p.e.}$ and $n^{\rm pass}_{\rm TP}=16$}. See Figure \ref{fig:gamma_aeffs} for a detailed description of the criteria used to generate each curve. \textit{Bottom panel:} Differential readout rates induced by diffuse proton events for different array-triggering scenarios. The \textit{black dashed} curve illustrates the single telescope readout rate, in the absence of any array trigger. The \textit{blue dotted} curve represents the rate at which two telescopes trigger coincidentally and initiate camera readout. \revision{}{The \textit{solid red} curve illustrates the collection area for events that are accepted by an array trigger implementing the parallax width algorithm with $P < 40\,{\rm m}${, assuming a super-pixel trigger threshold of 2.5$\,{\rm p.e.}$} and} reveals the additional vetoing of proton initiated events that is achieved. Each curve is annotated with the relative \textit{integral} readout rates, using the parallax width array trigger as the fiducial case. \textit{Both panels} correspond to a pass-through super-pixel multiplicity threshold $n^{\rm pass}_{\rm TP}=16$.}\label{fig:gamma-diff_and_proton_aeffs}
\end{figure*}

In addition to providing efficient rejection of unwanted background events, the parallax width algorithm must also retain a large fraction of genuine \gr\ events. The \textit{top} panel of Figure \ref{fig:gamma_aeffs}  illustrates the degree to which this is requirement is fulfilled by plotting  the energy-binned effective collection areas for the simulated array configuration, assuming a point-like \gr\ source, \textit{without} application of a pass-through for events yielding extensive high-multiplicity trigger images. {The parallax width algorithm requires at least two intersections (and therefore three triggering telescopes) to compute $P$. Air showers with a lower overall energy content produce less Cherenkov light and consequently, they trigger fewer telescopes. Accordingly, although the parallax-width algorithm operates effectively for images that are not minimally reconstructible, the effective collection area for a two-telescope, minimally reconstructible image criterion (\textit{green dashed} curves in Figure \ref{fig:gamma_aeffs} and the \textit{top} panel of Figure \ref{fig:gamma-diff_and_proton_aeffs}) nonetheless exceeds that of the parallax-width algorithm at energies below $\sim300\;\mathrm{GeV}$. However, it should be noted that the ability to exploit images that are not minimally reconstructible enables the parallax width algorithm to retain \textit{all} fully reconstructible three-telescope triggers at \textit{all} energies.}
The \textit{bottom} panel of Figure \ref{fig:gamma_aeffs}  demonstrates that adoption of a pass-through threshold  $n^{\rm pass}_{\rm TP} = 16$ further enhances the retention of minimally reconstructible \gr\ events by the DIAT. {We use \textit{minimally reconstructible} to describe those events that trigger \revision{}{at} least two telescopes and are subsequently found to yield at least two Cherenkov images that are useable for parameterization and geometric reconstruction of the corresponding air shower and the properties of its progenitor \gr.}

The \textit{top} panel of Figure \ref{fig:gamma-diff_and_proton_aeffs} uses simulated effective collection areas to demonstrate the response of the DIAT that implements the parallax width algorithm for \revision{an extended}{a diffuse} astrophysical \gr\ source, adopting $n^{\rm pass}_{\rm TP} = 16$. {Hereafter, we describe images that can be used for parameterization and geometric reconstruction of the corresponding air shower and the properties of its progenitor \gr\ as \textit{high quality}}. At low energies the rejection of events yielding $P>40\,{\rm m}$ retains a large fraction of events that yield three high-quality Cherenkov images, while at energies $\gtrsim1$ TeV, the pass-through trigger results in retention of \textit{all} minimally reconstructible events. The \textit{bottom} panel of Figure \ref{fig:gamma-diff_and_proton_aeffs} illustrates the power of $P$ to effectively reject cosmic-ray proton-initiated events. The \textit{blue} curve illustrates the {expected camera readout rate as a function of proton energy} for a traditional two-telescope multiplicity array trigger, that accepts events for which at least two neighbouring telescopes trigger. The parallax width trigger reduces the expected rate of incorrectly accepted CR events by a factor of $\sim4$, which is almost double the suppression that is achieved by the two-telescope multiplicity requirement. Moreover, the contrast in CR suppression efficacy is maximized at lower energies where the incident cosmic ray rate is largest.

\begin{figure*}[p]
 \centering
 \includegraphics[width=0.7\textwidth]{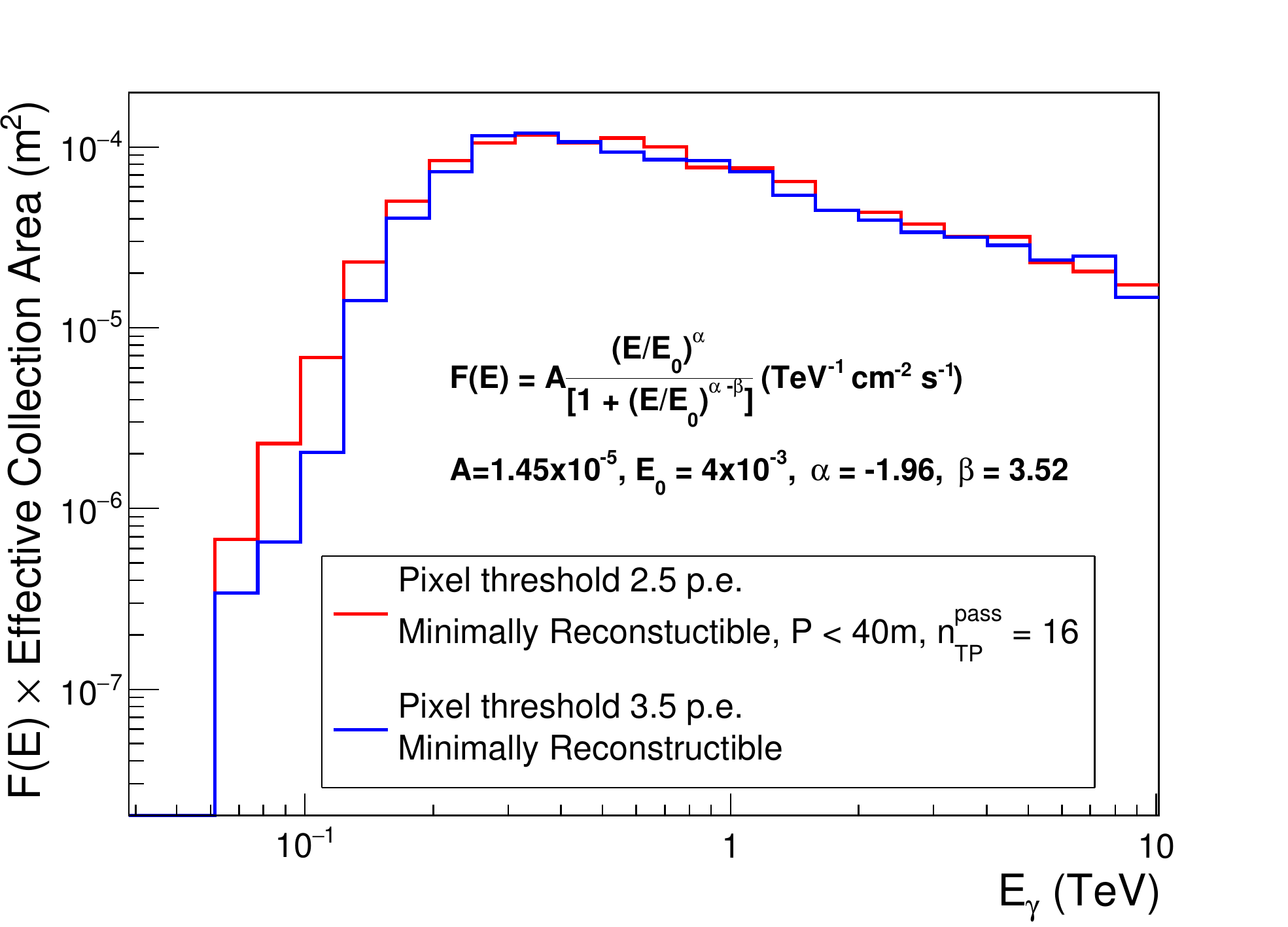}
 \includegraphics[width=0.7\textwidth]{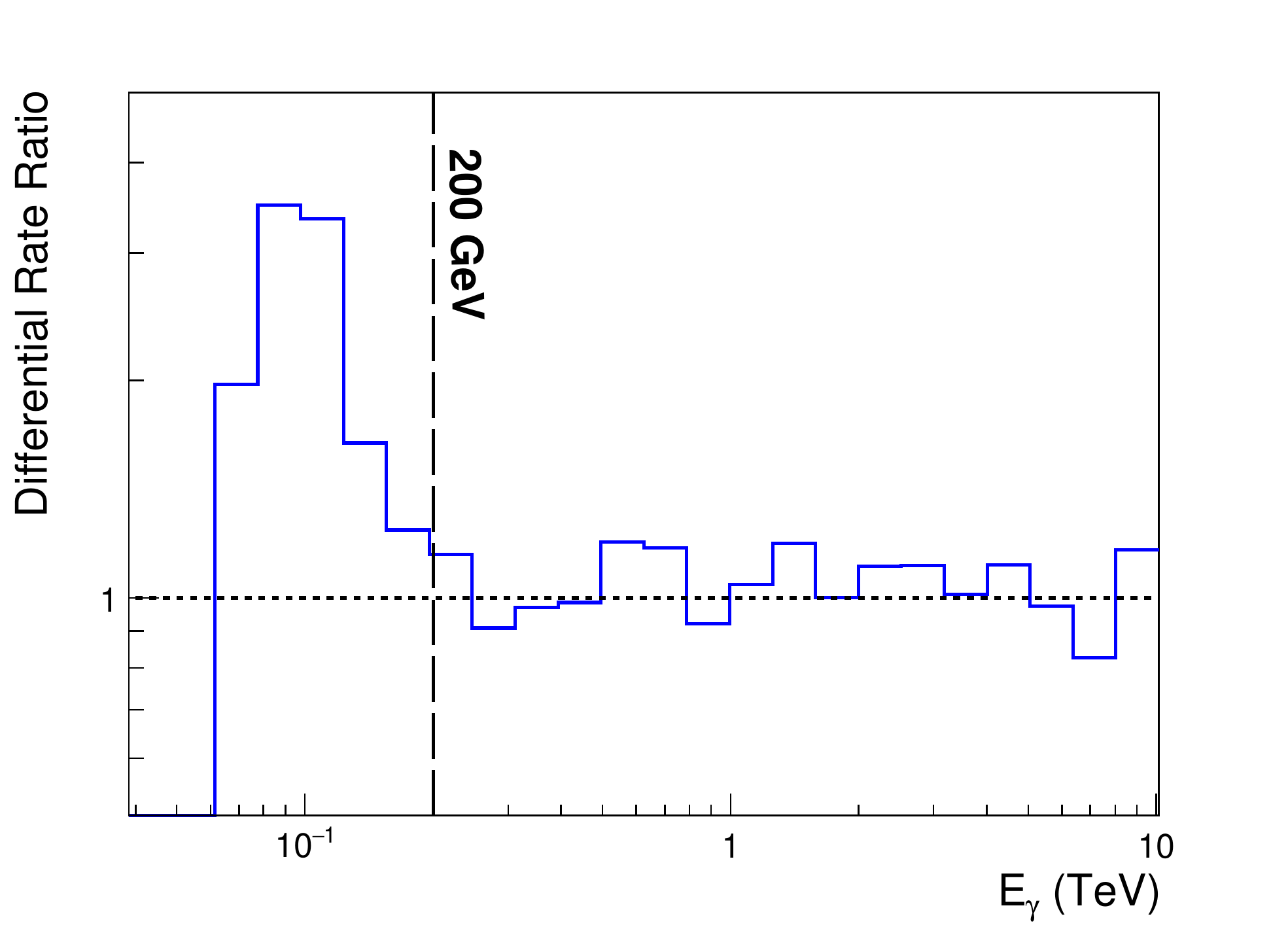}
 \caption{\small The expected array trigger rates for on-axis \gr-initiated events are shown in the \textit{top panel} for a Crab-pulsar-like spectral shape. The red curve corresponds to minimally reconstructible events that are retained after application of a trigger threshold corresponding to 2.5$\,{\rm p.e.}$ per super-pixel, and subsequent filtering using the parallax width algorithm with $P<40\,{\rm m}$, and a pass-through threshold $n^{\rm pass}_{\rm TP}=16$. For comparison, the \textit{blue} curve plots the expected array rate for minimally reconstructible events, assuming a higher super-pixel threshold of 3.5$\,{\rm p.e.}$, which is typically required to ensure stable operation under variable NSB illumination without an array trigger.
  The ratio of the \textit{red} and \textit{blue} curves is shown in the \textit{bottom} panel and illustrates that application of the parallax width discriminator enables effective recovery of low energy events by rendering a lower super-pixel threshold feasible.}\label{fig:event_recovery}
\end{figure*}

Figure \ref{fig:event_recovery} illustrates how a DIAT using the parallax width algorithm can be used to improve the low-energy sensitivity of the SCT subarray. Field-to-field variability of the NSB intensity between observations can induce unpredictable spikes in the array trigger rate if accidental coincidences between spurious single telescope triggers cannot be effectively suppressed. Typically, the required suppression is achieved by requiring a higher single-pixel trigger threshold, which increases the overall low-energy threshold of the array. The ability of the DIAT to perform real-time multi-telescope event vetoing allows lower pixel thresholds to be used while maintaining tractable array trigger rates. In Figure \ref{fig:event_recovery} (\textit{top panel}), the differential trigger rates for a Crab-pulsar-like \gr\ source
are compared for minimally reconstructible events that satisfy $P<40\,{\rm m}$ assuming a trigger pixel threshold of {2.5}$\,{\rm p.e.}$, with the set of minimally reconstructible events that
fulfil a traditional two-telescope multiplicity array trigger with $n_{\rm pe} = 3.5\,{\rm p.e.}$.
The lower pixel threshold made feasible by a hardware-level DIAT yields a substantial improvement in sensitivity below 200 GeV, achieving a factor of $\sim4$ enhancement in collection area at the lowest energies.


\section{Summary}\label{sec:conclusions}

Monte-Carlo simulations that model an array of SC Cherenkov Telescopes have been used to demonstrate that efficient rejection of cosmic-ray-initiated events is possible using an innovative, distributed, intelligent array trigger. The simulated trigger implements an algorithm (designated \textit{Parallax Width}) that performs hardware-level analysis using computed moments of the Cherenkov light distributions that are imaged by multiple telescopes. It successfully discriminates between background and genuine \gr\ triggers, while retaining a large majority of reconstructible events.

Simulated application of \revision{the}{} a DIAT that implements the \textit{Parallax Width} algorithm demonstrated several advantages over arrays that utilize traditional \textit{telescope multiplicity} triggers or operate without an array trigger.
\begin{itemize}
 \item By vetoing spurious single and multiple telescope triggers before data are read out, the algorithm reduces the array trigger rate by a factor of $\sim7$. This enables finely sampled events with heavy data payloads to be generated by participating SCTs without overwhelming the array data transfer infrastructure.
 \item Real-time consideration of data from multiple telescopes also allows the rate of NSB-induced array triggers to be controlled without increasing the super-pixel trigger thresholds. Indeed, a reduction in super-pixel trigger threshold corresponding to an additional photoelectron-equivalent count is rendered feasible.
 \item  The resultant enhancement of low-energy sensitivity increases the number of reconstructible events with energies between $\sim 100$ GeV and $\sim 200$ GeV, which may be particularly useful for studies of spectrally soft targets like the Crab pulsar.
\end{itemize}

The configuration of the simulated telescope array represents a realistic scenario but was \textit{not} tailored to maximize the efficacy of the parallax width trigger algorithm. Accordingly, a suitably optimized telescope array is likely to realize additional improvements in overall performance.

\section{Acknowledgements}
The authors thank Dr Michael Punch, Dr John Ward, and Dr Matthew Wood for the valuable advice and insight they provided with regard to this study. This paper has gone through internal review by the CTA Consortium.



\end{document}